\documentclass{scrartcl}

\usepackage[authoryear]{natbib}
\usepackage{amsmath, amssymb, amsthm}
\usepackage{todonotes}
\usepackage{units}
\usepackage{booktabs}
\usepackage[toc,page]{appendix}
\usepackage{subcaption}
\usepackage[ruled]{algorithm2e}

\usepackage[paperwidth=210mm, paperheight=297mm, margin = 20mm,footskip=.35in]{geometry}

\DeclareMathOperator{\E}{E}
\DeclareMathOperator{\var}{var}
\DeclareMathOperator{\GP}{GP}
\DeclareMathOperator{\predabil}{\eta}

\DeclareMathOperator{\vectorize}{vec}

\newcommand{\N}{\mathcal{N}}
\newcommand{\newobs}[1]{\tilde{#1}}
\renewcommand{\v}[1]{\mathbf{#1}}
\newcommand{\m}[1]{\mathbf{#1}}

\title{Modeling local predictive ability with power-transformed Gaussian process regression}
\author{Oscar Oelrich$^{a}$\thanks{Corresponding author: oscar.oelrich@stat.su.se. The computations in this chapter were enabled by resources provided by the National Supercomputer Centre (NSC), funded by Linköping University, and by resources in project SNIC 2020/15-262 provided by the Swedish National Infrastructure for Computing (SNIC) at UPPMAX, partially funded by the Swedish Research Council through grant agreement no. 2018-05973.} and\\ Mattias Villani$^{a}$}
\date{%
    $^a$Department of Statistics, Stockholm University\\%
}
\begin{document}
\maketitle

\abstract{A Gaussian process is proposed as a model for the posterior distribution of the local predictive ability of a model or expert, conditional on a vector of covariates, from historical predictions in the form of log predictive scores. Assuming Gaussian expert predictions and a Gaussian data generating process, a linear transformation of the predictive score follows a noncentral chi-squared distribution with one degree of freedom. Motivated by this we develop a noncentral chi-squared Gaussian process regression to flexibly model local predictive ability, with the posterior distribution of the latent GP function and kernel hyperparameters sampled by Hamiltonian Monte Carlo. We show that a cube-root transformation of the log scores is approximately Gaussian with homoscedastic variance, making it possible to estimate the model much faster by marginalizing the latent GP function analytically. A multi-output Gaussian process regression is also introduced to model the dependence in predictive ability between experts, both for inference and prediction purposes.  Linear pools based on learned local predictive ability are applied to predict daily bike usage in Washington DC.}

\section{Introduction}
\label{sec:1}

Combining predictions from several models or experts, be they point forecast \citep{bates_combination_1969} or forecast distributions \citep{yao_using_2018}, has a long history in statistics and econometrics; see \citet{wang_forecast_2023} for a recent review. A forecast combination can better quantify total uncertainty \citep{draper_assessment_1995} and improve accuracy \citep{wang_forecast_2023}. Some prominent example of aggregation methods include weighted averages with weights based on past forecast errors \citep{bates_combination_1969}, posterior model probabilities \citep{hoeting_bayesian_1999} and historical performance of the experts' density forecasts \citep{hall_combining_2007, geweke_optimal_2011}. 

The out-of-sample predictive accuracy of an expert, which could be a formal model or a human, is typically treated as a fixed quantity to be estimated \citep*{gelman_understanding_2014}. Our aim here is to model \emph{local predictive ability}, i.e. predictive ability as a function of a vector of covariates \(\mathbf{z}\), to capture how the predictive abilities of a set of experts vary over the space of \(\mathbf{z}\) \citep{yao_bayesian_2021, li_bayesian_2023, oelrich_local_2024}. We aim to obtain the posterior distribution of the local predictive ability to properly quantify the uncertainty of the expert's predictive ability.

Since one use of measures of predictive ability is to combine, or pool, expert predictions, we call the variables in \(\mathbf{z}\) \emph{pooling variables} and the space of \(\mathbf{z}\) the \emph{pooling space} \citep{oelrich_local_2024}. The choice of pooling variables is determined by the \emph{decision maker} \citep{lindley_reconciliation_1979} making the aggregated prediction, based on what she believes are important determinants of the experts' local predictive abilities. In an economic application a measure of the business cycle or a survey of expectations could for example be a pooling variable, if the forecasting experts are expected to have different forecasting accuracy depending on where the economy is in the business cycle. 

Our focus is on the accuracy of an expert's forecast  \emph{density} \citep{gelman_understanding_2014}. Density forecasts are typically part of the output from probabilistic models, but are also increasingly reported from human experts. For example, the histogram densities for key macroeconomic variables from the Survey of Professional Forecasters conducted by the Federal Reserve Bank of Philadelphia for the US \citep{croushore_fifty_2019} and the European Central Bank for the Euro Area \citep{de2019twenty} or probabilistic forecasts of temperature and precipitation in meteorology from a combination of subjective experts, historical data and complex deterministic scientific models \citep{council2008enhancing}.
 
A theoretically motivated measure of predictive ability is the expected log predictive density (ELPD), which can be unbiasedly estimated by average log predictive density evaluations (log scores) on a test dataset. The distribution of the log scores is typically complex however \citep{sivula_unbiased_2022}, making inferences for ELPD challenging. The likelihood of the observed log scores depends not only on the model/expert but also on the data-generating process (DGP). For example, using a simple model with additive noise will typically work poorly, as when the predictive distribution and the DGP are normal, the log scores will follow a scaled and translated noncentral \(\chi^2_1\) distribution \citep{sivula_uncertainty_2022}. 
 
Further, specifying the relationship between the set of covariates and predictive ability is a challenging problem in itself. Certain variables may be expected to be important for predictive ability but the exact functional form may be harder to pinpoint. In a situation of little prior information it makes sense for the decision maker to use a non-parametric model that can capture variations in predictive ability over an arbitrary pooling space in a flexible manner. \citet{oelrich_local_2024} propose the non-parametric caliper estimator of local predictive ability, which averages log scores locally in pooling space. The caliper estimator is flexible, but has the potential drawback that the inferred local ability surface is discontinuous, and the lack of smoothness may result in large variability for the local estimates. 

We here propose a flexible noncentral \(\chi^2\) Gaussian process (GP) regression, where the latent GP surface models the conditional ELPD over the space of pooling variables in \(\mathbf{z}\). A Gaussian process explicitly models the correlation properties, but makes otherwise minimal assumptions about the specific relationship between the covariates and the ELPD. The smoothness and other hyperparameters in the covariance kernel are assumed unknown and are inferred jointly with the latent GP surface. A Gaussian process model for local ELPD is attractive since it gives a direct quantification of the uncertainty about the underlying ELPD surface for each expert.

Separate GPs can be estimated for each expert, implicitly assuming independence when forecasts are pooled based on local predictive ability.
However, correctly capturing any dependence structure between experts is useful both from a pure inference standpoint, as it allows us to better understand the relationship between the predictions of the experts, as well as for creating better aggregate predictions \citep{mcalinn_multivariate_2020}.
The latter point is clearly illustrated by \citet{winkler_combining_1981}, who shows that an expert with relatively low predictive ability can nevertheless get a substantial weight in a combined forecast if the expert's forecast error is negatively correlated with the forecast errors of the other experts. We therefore relax the independence assumption by also proposing a multi-output GP to \emph{jointly} model the local predictive ability of a set of \(K\) experts over the pooling space.

The multi-output GP uses a kernel function that is constructed by correlating a set of underlying independent GPs, one for each expert.
This construction makes it possible to use separate smoothness kernel parameters, or even completely different classes of kernels, for each expert.
It also gives us access to interpretable parameters that describe the correlation between experts.  
The multi-output model allows us to jointly infer the latent GP surfaces of all experts, which gives a more full quantification of the uncertainty of the ELPD surface of prediction pools based on the set of experts.

The joint posterior distribution of the conditional ELPD and kernel hyperparameters are sampled by Hamiltonian Monte Carlo (HMC) using Stan \citep{stan_development_team_stan_2024}. The dimension of the sampled conditional ELPD grows with the sample size however, and inferences are time-consuming even for moderately large datasets. We show empirically that a cube-root transformation of the log scores is approximately Gaussian with homoscedastic variance. This makes it possible to integrate out the latent GP surface analytically and sample the low-dimensional marginal posterior kernel hyperparameters by HMC, which is much faster and scalable for larger datasets. We show that the ELPD surface can then be easily sampled by standard GP methods \citep[Chapter~2]{rasmussen_gaussian_2006} combined with analytically available third moments of Gaussian processes. In the more general case when either the expert or the data generating process is non-Gaussian we propose to use a general power transformation of the ELPD and to estimate the power transformation using HMC jointly with the kernel hyperparameters.

The chapter proceeds as follows. Section \ref{sec_framework} defines local predictive ability and describes the noncentral \(\chi^2\) Gaussian process (GP) regression for estimating it, along with a computationally more efficient single- and multi-output GP regression on power-transformed log scores. Section \ref{sec_empirical} consists of an empirical example analyzing local predictive ability for a set of models in a bike sharing dataset, along with several methods to pool predictive distributions based on estimated predictive ability. Section \ref{sec_conclusions} concludes.

\section{A framework for estimating local predictive ability}\label{sec_framework}

This section defines local predictive ability within a decision maker framework \citep*{oelrich_local_2024}, discusses the implications of using log scores as data to estimate local predictive ability, and proposes Gaussian process regression as a model for estimating both joint and marginal local predictive ability.

\subsection{Local predictive ability}

We consider the problem where a decision maker is presented with a set of \(K\) experts, either in the form of models or opinionated humans, and wants to estimate how well the experts make predictions, based on their historical performance. We assume that the decision maker has access to historic predictions made by the experts in the form of log predictive density scores $\ell_{ik} = \log p_k(y_i \mid \mathbf{y}_k)$, assessing the prediction of observation $y_i$ in the $k$th expert's predictive density given a set of training data $\mathbf{y}_k$.
While nothing prevents us from considering other measures, such as mean squared error, the derived model for predictive ability depends on the chosen measure. The log predictive density score is considered the gold standard for measuring predictive performance since it has the unique advantage of being both local and proper \citep*{gneiting_strictly_2007}, and is therefore commonly used in model selection. Further, the expected log predictive density of a predictive distribution is proportional to the Kullback-Leibler divergence with regards to the data-generating process \citep*{hall_combining_2007}.

The decision maker uses the expected log predictive density for a new data point \(\newobs{y}\) from the data-generating process \(p_{*}(\cdot)\) to define the predictive ability of expert \(k\)
\begin{equation}\label{elpd}
    \eta_k
    \equiv
    \E_{p_{*}(\newobs{y})}
    \left[
        \log p_k(\newobs{y}\mid \mathbf{y}_k)
    \right] =
    \int_{-\infty}^{\infty} 
    \left[
        \log p_k(\newobs{y}\mid \mathbf{y})
    \right]
    p_{*}(\newobs{y})\,d\newobs{y},
    \end{equation}
where \(\newobs{y}\) is a new (out-of sample) observation, \(p_{*}(\cdot)\) is the density of the data-generating process, and \(\mathbf{y}_k\) is the training data the expert has access to. The variable of interest can be univariate or multivariate, but we will for simplicity assume that it is univariate here.

The decision maker further believes that the predictive ability of the experts vary over a set of pooling variables $\mathbf{z}$ \citep{oelrich_local_2024}.
She is therefore not only interested in determining how good the experts are at making predictions in general, but how good they are at making predictions conditional on the pooling variables, \(\mathbf{z}\). We call this conditional predictive ability the \emph{local predictive ability} of the expert at \(\mathbf{z}\)
\begin{equation}\label{lelpd}
    \eta_k(\mathbf{z}) \equiv
    \E_{p_{*}(\newobs{y}\vert \mathbf{z})}
    \left[
        \log p(\newobs{y}\mid \mathbf{y}_k, \mathbf{z})
    \right]
     =
    \int_{-\infty}^{\infty} 
    \left[
        \log p(\newobs{y}\mid \mathbf{y}_k, \mathbf{z})
    \right]
     p_{*}(\newobs{y}\vert \mathbf{z})\,d\newobs{y},
\end{equation}
where the data generating process $p_{*}(\newobs{y}\vert \mathbf{z})$ may or may not depend on the pooling variables $\mathbf{z}$. 

\subsection{Distribution of the log scores}
The log scores of the experts, $\ell_{ik}$ for $k=1, \ldots,K$ and $i=1, \ldots,n$, are treated as data by the decision maker \citep*{lindley_reconciliation_1979} from which she infers how the ELPD of each expert varies over the pooling space. The choice of pooling variables is therefore a pure modeling problem from the viewpoint of the decision maker. The decision maker does not need to know if the experts have access to the pooling variables when making their predictions, but this information can of course affect the modeling choices of the decision maker. 

The distribution of the local predictive ability is determined by the predictive distribution of the expert, which is typically known, and the data-generating process, which is typically unknown.
In order to specify a parametric model for the log scores, we therefore need access to the predictive distribution of the expert. Different predictive distribution will lead to different models of local predictive ability.

\begin{figure}[htbp]
    \centering
    \includegraphics[width = 12cm]{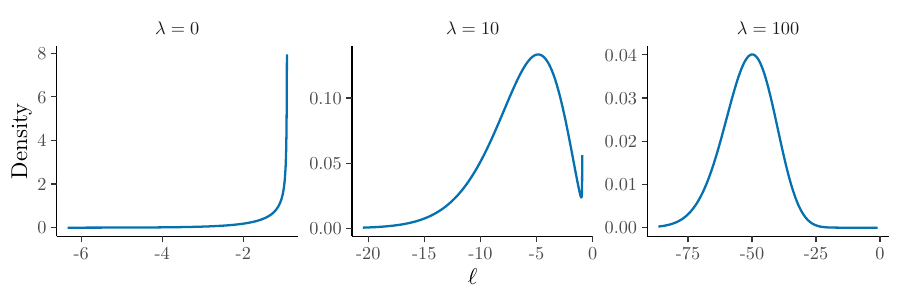}
    \caption{Distribution of log scores for a Gaussian expert and DGP with equal variance for three different $\lambda_k=(\mu_{*}-\mu_{k})^2$.}
    \label{fig_logscores_dist}
\end{figure}

Consider the practically important scenario where the expert predictive distributions, as well as the data generating process, are all Gaussian. Specifically, let the predictive distribution of expert \(k\) for the data point \(y_i\) be \({y_i\mid \mathbf{y}_k \sim \N(\mu_{ik}, \sigma_{ik}^2)}\), where $\mathbf{y}_k$ is the training data used by the $k$th expert. The log score is then 
\begin{equation}\label{logscores_dist}
    \ell_{ik} \equiv \log p_k(y_i \mid \mathbf{y}_k) = 
            -\frac{1}{2}\log(2\pi\sigma_{ik}^2)
            -\frac{1}{2\sigma_{ik}^2}(y_i - \mu_{ik})^2\;.
\end{equation}
If we let the data-generating process be given by \({y_i \sim \N(\mu_{i*}, \sigma_{i*}^2)}\), we can trivially rewrite Eq. \eqref{logscores_dist} as
\begin{equation}
    \ell_{ik} \equiv \log p_k(y_i \mid \mathbf{y}_k) = 
            -\frac{1}{2}\log(2\pi\sigma_{ik}^2)
            -\frac{\sigma_{i*}^2}{2\sigma_{ik}^2}\left(\frac{y_i - \mu_{ik}}{\sigma_{i*}}\right)^2.
\end{equation}
Since \((y_i - \mu_{ik})/\sigma_{i*} \sim \N(\mu_{i*}-\mu_{ik}, 1)\), it follows that \citep{sivula_uncertainty_2022}
\begin{equation}\label{eq:nc_lincomb}
    \ell_{ik} \overset{d}{=} a_{ik} - b_{ik}\cdot X_{ik},
\end{equation}
where $X_{ik} \sim \chi^2_1(\lambda_{ik})$ with noncentrality parameter $\lambda_{ik}=(\mu_{i*}-\mu_{ik})^2/\sigma_{i*}^2$, $a_{ik}=-\frac{1}{2}\log(2\pi\sigma_{ik}^2)$ and $b_{ik}=\sigma_{i*}^2/(2\sigma_{ik}^2)$. Hence, the noncentrality parameter \(\lambda_{ik}\) depends on the difference between the predictive mean of the expert and the mean of the data-generating process; the shape of the distribution of the log scores therefore varies dramatically with $\mu_{i*}-\mu_{ik}$ and is typically highly non-Gaussian for all \(b\), see Fig. \ref{fig_logscores_dist} for an illustration. 

Since $a_{ik} = -\frac{1}{2}\log(2\pi\sigma_{ik}^2)$ only depends on the expert's predictive distribution and is therefore assumed known, we can define $\ell'_{ik} \equiv -(\ell_{ik}-a_{ik})$ and write the model in \eqref{eq:nc_lincomb} as
\begin{equation}\label{eq:modelwithdifferent_bik}
\ell'_{ik} \mid \lambda_{ik},b_{ik} \sim \chi^2_1(\lambda_{ik}, b_{ik}),
\end{equation}
where $\chi^2_1(\lambda, b)$ denotes a \emph{scaled} noncentral $\chi^2$ with one degree of freedom, noncentrality parameter $\lambda$ and scale parameter $b$. A useful simplification of the model in \eqref{eq:modelwithdifferent_bik} is to assume that $b_{ik} = b_k$ for all $i = 1,\ldots,n$, i.e. that the variance ratio $\sigma_{i*}^2/\sigma_{ik}^2$ is constant over observations for a given expert. 


\subsection{Scaled noncentral \(\chi^2\) Gaussian process regression}

The setup with Gaussian data generating process and experts in the previous subsection suggests a scaled noncentral \(\chi^2\) regression for modeling the local predictive ability of experts. We will here propose a nonparametric Bayesian model where the log of the noncentrality parameter is modeled by a Gaussian process over the pooling space.

A Gaussian process (GP) generalizes the Gaussian distribution from a random variable to a stochastic process \citep*{rasmussen_gaussian_2006}.
In simple terms, a Gaussian process is an infinite sequence of function values with different inputs \(\{f(\mathbf{z}_1), f(\mathbf{z}_2), f(\mathbf{z}_3),\dots\}\), any subset of which follows a multivariate Gaussian distribution.
Each GP has a covariance function, or kernel, which depends on the input values and specifies how the covariance matrix for each subset of the sequence should be constructed.
The specific covariance function we select determines our prior regarding how smoothly the outputs vary over the input space.
The covariance function we use in the examples and applications in this chapter is the squared exponential with automatic relevance determination (ARD) \citep{neal_bayesian_1996}

\begin{equation}\label{squared_exponential}
    k_{\operatorname{SE}}(\mathbf{z}_i, \mathbf{z}_j) =
    \alpha^2 \exp \left(
        -\frac{1}{2} 
        (\mathbf{z}_i - \mathbf{z}_j)
        (\operatorname{diag}(\boldsymbol{l})^{-2})
        (\mathbf{z}_i - \mathbf{z}_j)
    \right) + \delta_{ij}\sigma_n^2,
\end{equation}
where \(\boldsymbol{l}\) is a vector of \emph{length scales}, \(\alpha\) denotes the signal standard deviation, \(\sigma_n^2\) denotes the noise variance, and \(\delta_{ij}\) is the Kronecker function, which takes the value \(1\) when \(i=j\) and \(0\) otherwise.
Kernels where the length scales of different inputs are allowed to differ are referred to as automatic relevance determination kernels, as the length scales are inversely related to relevance \citep{rasmussen_gaussian_2006}.

Our noncentral \(\chi^2\) Gaussian process model for the log scores is
\begin{align}\label{eq:noncentralGP}
\begin{split}
    &\ell'_{ik} \mid \lambda_{k}(\mathbf{z}_i),b_{k} \overset{\mathrm{indep.}}{\sim} \chi^2_1(\lambda_{k}(\mathbf{z}_i), b_{k}) \\
    &\log{\lambda_{k}(\mathbf{z})} \sim
    \GP\big(\mu_k(\mathbf{z}), g_k(\mathbf{z}, \mathbf{z}^\prime)\big) \\
    &b_k \sim \N^+(1/2, \psi_b^2),
\end{split}
\end{align}
where \(\N^+(\cdot, \cdot)\) denotes a normal distribution truncated at $0$. Note that \(b_k>0\), and \(b_k=1/2\) when the variance of the DGP and the expert match. Further, note that the GP is on the log of the noncentrality parameter, ensuring that \(\lambda_{k}(\mathbf{z})\) stays non-negative. Since the mean of a \(\chi^2_1(\lambda,b)\) variable is $b(1+\lambda)$, the Gaussian process in \eqref{eq:noncentralGP} directly implies a flexible prior for the mean function of the log scores, i.e. the local ELPD. We will refer to the model in \eqref{eq:noncentralGP} as \(\GP(\chi^2_1)\). The mean function $\mu_k(\cdot)$ and the covariance kernel $g_k(\cdot,\cdot)$ are subscripted by $k$ since their hyperparameters are allowed to be different across the $K$ experts. Any valid kernel function can be used. For notational simplicity, we let $\boldsymbol{\theta}_k$ be the vector with all the hyperparameters in the mean and kernel function of the GP prior.   

The ultimate goal of the decision maker is to compute the posterior distribution of the local ELPD $\predabil_k(\newobs{\mathbf{z}}) = \E_{p_{*}}(\tilde{\ell}_{k}) =  \tilde a_{k} - b_k(1 + \lambda_{k}(\tilde{\mathbf{z}}))$ at a new point $\tilde{\mathbf{z}}$ in pooling space for each expert. This is achieved by sampling from the joint posterior 
\begin{equation}\label{eq:fulljointposterior}
    p\big(\lambda_{k}(\tilde{\mathbf{z}}),\boldsymbol{\lambda}_{k},b_k,\boldsymbol{\theta}_k \mid \boldsymbol{\ell}^\prime_{k}\big),
\end{equation}
where $\boldsymbol{\ell}^\prime_{k}=(\ell^\prime_{1k},\ldots,\ell^\prime_{nk})^\top$ and \( \boldsymbol{\lambda}_{k} = (\lambda_{k}(\mathbf{z}_1),\ldots,\lambda_{k}(\mathbf{z}_n))^\top \) is the vector of noncentrality parameters in the training data.
We sample the potentially high-dimensional posterior in \eqref{eq:fulljointposterior} using Hamiltonian Monte Carlo (HMC) in Stan  \citep{stan_development_team_stan_2024, gabry_cmdstanr_2022}.

\subsection{Cube root transformed Gaussian process regression}
\label{GP_one_third}

HMC sampling for the scaled noncentral \(\chi^2\) Gaussian process regression is efficient, but necessarily slow in even moderately large datasets since \(\boldsymbol{\lambda}_{k}\) is high-dimensional. This subsection presents a cube root transformation of the log scores which makes the log scores approximately normal with constant variance. This allows us to analytically integrate out the high-dimensional latent ELPD surfaces from the posterior, thereby substantially speeding up the computation of the ELPD posterior at a new point in the pooling space. We will refer to this model in tables and figures as \(\GP\left(\nicefrac{1}{3}\right)\).

\begin{figure}[htbp]
    \centering
    \includegraphics[width=12cm]{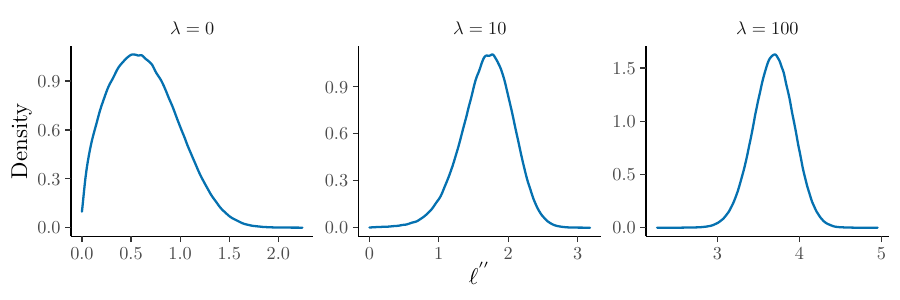}
    \caption{Approximate distribution of transformed log scores for a Gaussian expert and DGP with equal variance for three different $\lambda_k=(\mu_{*}-\mu_{k})^2$.}
    \label{fig_logscores_trans}
\end{figure}

\begin{figure}[htbp]
    \centering
    \includegraphics[width=12cm]{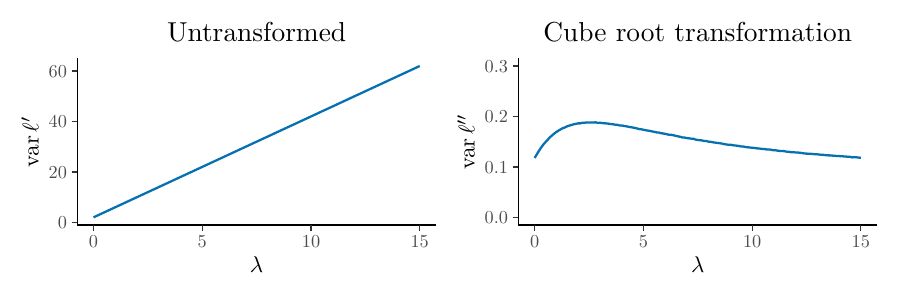}
    \caption{Variance of log scores as function of $\lambda=(\mu_{*}-\mu_{k})^2$ for a Gaussian expert and DGP with equal variance.}
    \label{fig_logscores_trans_var}
\end{figure}

Starting from the linear transformation $\ell'_{ik} = -(\ell_{ik}-a_{ik})$, we further transform the log scores \(\ell_{ik}^{\prime}\) by taking the cube root, which is motivated by \citet{abdel-aty_approximate_1954} who shows that a version of the Wilson-Hilferty transformation of the \(\chi^2_v\) distribution approximately holds for \(X \sim \chi_v^2(\lambda)\) as well. That is 
\begin{equation}
    \left(  
        \frac{X}{v+\lambda}
    \right)^{1/3}
    \dot\sim \,\N,
\end{equation}
where the mean and variance of the normal distribution depends on the noncentrality parameter \(\lambda\) as well as the degrees of freedom. Since the noncentrality parameter $\lambda$ is not known, we use a plain cube-root $X^{1/3}$ transformation when transforming the log scores. Figure \ref{fig_logscores_trans} shows that the this transformation makes the log scores approximately Gaussian; compare with Fig. \ref{fig_logscores_dist}. As \(\lambda\) goes towards infinity the variance goes to zero, but Fig. \ref{fig_logscores_trans_var} shows that the variance is approximately constant in a range of reasonable \(\lambda\) values. We use the notation \(\ell'' = (\ell')^{1/3}\) for these twice transformed log scores.

Once our log scores have been transformed to approximate normality with constant variance, we can use a Gaussian process with homoscedastic Gaussian noise to model the predictive ability of each expert: 
\begin{align}
\begin{split}
    &\ell''_{ik} \mid f_{k}(\mathbf{z}_i), \sigma_{k} \overset{\mathrm{indep.}}{\sim} \N(f_{k}(\mathbf{z}_i), \sigma^2_{k}) \\
    &f_{k}(\mathbf{z}) \sim
    \GP\big(\mu_k(\mathbf{z}), g_k(\mathbf{z}, \mathbf{z}^\prime)\big) \\
    &\sigma_k^2 \sim \N^+(0, \psi_{\sigma}^2)\;.
\end{split}
\label{cubeGP_model}
\end{align}
Again, let $\boldsymbol{\theta}_k$ denote the vector of hyperparameters in the mean and kernel functions of the GP for the $k$th expert.

Since $\ell_{ik} = a_{ik} - \left(\ell''_{ik}\right)^3$, the ELPD at the test point $\newobs{\mathbf{z}}$ is $\eta_k(\newobs{\mathbf{z}}) = \tilde{a}_{k} - \E_{p_{*}}\left[\left(\tilde{\ell}''_{k}\right)^3\right]$. For a general \(X \sim \N(\mu, \sigma^2)\) we have
\begin{equation}
    \E(X^3) = \mu^3 + 3 \mu \sigma^2,
\end{equation}
and hence
\begin{align}
\begin{split}
    \E_{p_{*}}\big[(\tilde{\ell}''_{k})^3\big] &= [\E_{p_{*}}(\tilde{\ell}''_{k})]^3 + 3\E_{p_{*}}(\tilde{\ell}''_{*k}) \cdot \var_{p_{*}}(\tilde{\ell}''_{k}) \\
    &= f_k^3(\tilde{\mathbf{z}}) + 3f_k(\tilde{\mathbf{z}})\sigma_k^2 \;.
\end{split}
\end{align}
We therefore have the following expression for the local ELPD at $\tilde{\mathbf{z}}$
\begin{equation}\label{eq:ELPDcuberoot}
    \eta_k(\tilde{\mathbf{z}}) = \tilde{a}_{k} - f_k^3(\tilde{\mathbf{z}}) - 3f_k(\tilde{\mathbf{z}})\sigma_k^2\;.
\end{equation}

To obtain the posterior for $\eta_k(\newobs{\mathbf{z}})$, we need the joint posterior $p\big(f_k(\newobs{\mathbf{z}}),\sigma_k, \boldsymbol{\theta}_k\mid \boldsymbol{\ell}''_{k}\big)$. Due to the approximate normality induced by the cube root transformation, we can analytically integrate out $\boldsymbol{f}_k = (f_k(\mathbf{z}_1),\ldots,f_k(\mathbf{z}_n))^\top$ in the training data, and obtain the marginal posterior of $\sigma_k$ and the kernel hyperparameters $\boldsymbol{\theta}_k$ in closed form \citep[Eq.~2.28]{rasmussen_gaussian_2006}
\begin{equation}\label{eq:marglikegaussianGPreg}
    p\big(\sigma_k,\boldsymbol{\theta}_k \mid \boldsymbol{\ell}''_{k}\big) = \int p\big(\boldsymbol{f}_{k},\sigma_k,\boldsymbol{\theta}_k \mid \boldsymbol{\ell}''_{k}\big)d\boldsymbol{f}_{k},
\end{equation}
where $\boldsymbol{\ell}_k''$ is a vector with transformed log scores in the training data. This low-dimensional marginal posterior can be sampled with Hamiltonian Monte Carlo (HMC) using Stan \citep{stan_development_team_stan_2024, gabry_cmdstanr_2022} in a fraction of the time that it takes to sample with full joint posterior for the $\mathrm{GP}(\chi_1^2)$ model in \eqref{eq:fulljointposterior}. 

For each HMC sampled $(\sigma_k,\boldsymbol{\theta}_k)$ we can now easily sample $f_k(\tilde{\mathbf{z}})$ from \citep*{rasmussen_gaussian_2006}
\begin{equation}\label{eq:predictivefstar}
    f_k(\tilde{\mathbf{z}})
    \mid 
    \boldsymbol{\ell}_k'', \mathbf{Z}, \tilde{\mathbf{z}}, \sigma_k, \boldsymbol{\theta}_k 
    \sim 
    \N\left[
        \bar{f}(\tilde{\mathbf{z}}), \operatorname{var}(f(\tilde{\mathbf{z}})) 
    \right],
\end{equation}
where $\mathbf{Z}$ is the matrix with pooling variables in the training data, and
\begin{align*}
    \bar{f}(\tilde{\mathbf{z}}) &= \mu_k(\newobs{\mathbf z}) +
    \operatorname{G}(\tilde{\mathbf{z}}, \mathbf{Z})
    \left[
        \operatorname{G}(\mathbf{Z}, \mathbf{Z}) + \sigma_k^2 I
    \right]^{-1} (\boldsymbol{\ell}_k'' - \boldsymbol{\mu}_k),\\
    \operatorname{var}(f(\tilde{\mathbf{z}})) &=
    \operatorname{G}(\tilde{\mathbf{z}}, \tilde{\mathbf{z}}) -
    \operatorname{G}(\tilde{\mathbf{z}}, \mathbf{Z})
    \left[
        \operatorname{G}(\mathbf{Z}, \mathbf{Z}) + \sigma_k^2 I
    \right]^{-1}
    \operatorname{G}(\mathbf{Z}, \tilde{\mathbf{z}}),
\end{align*}
where, for example, $G(\mathbf{Z}, \mathbf{Z})$ is the $n\times n$ matrix with covariances among the $f(\mathbf{z})$ in the training data given by the kernel function $g_k(\cdot,\cdot)$. The posterior draws of $\sigma_k$ from \eqref{eq:marglikegaussianGPreg} and $f_k(\tilde{\mathbf{z}})$ from \eqref{eq:predictivefstar} can finally be inserted into \eqref{eq:ELPDcuberoot} to obtain draws from the posterior of the local ELPD $\eta_k(\tilde{\mathbf{z}})$. The procedure is summarized in Algorithm \ref{alg:GPcuberoot}.

\begin{algorithm}
    \caption{Posterior sampling for the ELPD at $\tilde{\mathbf{z}}$ in the \(\GP\left(\nicefrac{1}{3}\right)\) model}\label{alg:GPcuberoot}
    \KwIn{Number of posterior draws $M$, vector of pooling variables $\tilde{\mathbf{z}}$ and $\tilde a_{k}=-(1/2)\log(2\pi \tilde \sigma_k^2)$ for expert $k$.
    }
    \vspace{2mm}
  
  Generate $M$ draws $\{\sigma_k^{(j)}, \boldsymbol{\theta}_k^{(j)}\}_{j=1}^M$ from $p\big(\sigma_k,\boldsymbol{\theta}_k \mid \boldsymbol{\ell}''_{k}\big)$ in \eqref{eq:marglikegaussianGPreg} using HMC.
  \vspace{1mm} \\
  \For {$j = 1,\dots, M$}{
  Sample from the posterior predictive $f_k^{(j)}(\tilde{\mathbf{z}}) \mid 
    \boldsymbol{\ell}_k'', \mathbf{Z}, \tilde{\mathbf{z}}, \sigma_k^{(j)}, \boldsymbol{\theta}_k^{(j)}$ in \eqref{eq:predictivefstar}\\
    Compute the ELPD draw: $\eta_k^{(j)}(\tilde{\mathbf{z}}) = \tilde a_{k} - (f_k^{(j)}(\tilde{\mathbf{z}}))^3 - 3f_k^{(j)}(\tilde{\mathbf{z}})(\sigma_k^{(j)})^2$
  \\ \vspace{1mm}
  }
  \vspace{1mm}
 \KwOut{Sample from ELPD posterior: $\eta_k^{(1)}(\tilde{\mathbf{z}}), \eta_k^{(2)}(\tilde{\mathbf{z}}), \ldots, \eta_k^{(M)}(\tilde{\mathbf{z}})\;.$}
\end{algorithm}

As the \(\GP(\nicefrac{1}{3})\) adds an approximation layer on top of the original model, we want to ensure that the gains in computational speed do not come at the cost of a drop in performance. To this end, we conduct a simulation study that shows good performance of the \(\GP(\nicefrac{1}{3})\) model compared to the \(\GP(\chi^2_1)\) model. For details, see Appendix.

\subsection{General power-transformed Gaussian process regression}

The cube root transformation is suitable when the log scores follow a scaled noncentral $\chi^2_1$ distribution, which is the correct model in the important case when both the DGP and the expert is Gaussian. Deviations from Gaussianity lead to new models that need to be derived on a case by case basis. The posterior distribution of the ELPD can be sampled with HMC as in the noncentral $\chi^2_1$ model described above. There may be new transformations that convert the log scores to something that can be modeled by a Gaussian process regression with Gaussian errors, but also this needs to be determined for each particular model.  

Moderate departures from Gaussianity can however be handled with a more general power transformation than the cube root transformation. Let $\ell^{(\alpha)}_{ik} = (\ell_{ik}')^\alpha$ for $\alpha>0$ be a general power transformation. The ELPD will then be $\eta_k(\tilde{\mathbf{z}}) = \tilde{a}_{k} - \E_{p_{*}}\big[(\tilde{\ell}^{(\alpha)}_{k})^{1/\alpha}\big]$. The fractional moments $\E_{p_{*}}\big[(\tilde{\ell}^{(\alpha)}_{k})^{1/\alpha}\big]$ can again be expressed in terms of $f_k(\tilde{\mathbf{z}})$ and $\sigma_k^2$ as in \eqref{eq:ELPDcuberoot} using the formulas in \citet{haldane_moments_1942}. The power parameter, $\alpha$, can be sampled in the HMC step along with $\sigma_k$ and the kernel hyperparameters in $\boldsymbol{\theta}_k$. The prior for $\alpha$ can be centered on $\alpha=1/3$, i.e. the cube root transformation.

\subsection{The multi-output \(\operatorname{GP}(\nicefrac{1}{3})\) model}

The \(\operatorname{GP}(\nicefrac{1}{3})\) model proposed in Sect. \ref{GP_one_third} can be used to model the predictive ability of each of a set of experts. If these estimates are then used to create a pooled predictive distribution two independence assumptions are implied: independence in the predictive abilities $\eta_1(\mathbf{z}_i), \ldots, \eta_K(\mathbf{z}_i)$, and independence in the errors, $\ell''_{ik} - f_{k}(\mathbf{z}_i)$ for $i=1,\ldots,n$, conditional on the predictive ability. In this section, we propose a generalisation of the \(\operatorname{GP}(\nicefrac{1}{3})\) model to a multi-output GP that relaxes both of these assumptions.

Let \(\ell_{ik}^{\prime\prime} = \left(a_{ik} - \ell_{ik}\right)^{\nicefrac{1}{3}}\) denote the cube-root transformed log scores. The distribution of \(\ell_{ik}^{\prime\prime}\) for \(K\) experts jointly, \(\boldsymbol{\ell}_{i}^{\prime\prime} = (\ell_{i1}^{\prime\prime},\ldots,\ell_{iK}^{\prime\prime})^\top\), can then be modeled using a multi-output GP with Gaussian noise
\begin{align}
\begin{split}
        \boldsymbol{\ell}''_{i} \mid 
        \v{f}(\v{z}_i), 
        \boldsymbol{\Sigma}
       &\sim
        \N\left(\v{f}(\mathbf{z}_i), \boldsymbol{\Sigma}\right)\\
        \mathbf{f}(\v{z}) \mid \boldsymbol{\Theta}, \m{C}
        &\sim
        \operatorname{GP}\left(\boldsymbol{\mu}(\v{z}), g(\v{z}, \v{z}^\prime)\right),
\end{split}
\label{multioutGP}
\end{align}
where \(\v{f}(\v z) = (f_1(\v z),\ldots,f_K(\v z))^\top\) is a multi-output GP with hyperparameters \(\boldsymbol{\Theta} = \{\boldsymbol{\theta}_1\),\ldots,\(\boldsymbol{\theta}_K\}\) including a vector of parameters \(\boldsymbol{\theta}_k\) for the mean and covariance function of each expert as well, and \(\m{C}\) is a \(K \times K\) matrix that will be used below to correlate the experts' abilities \(\mathbf{f}\). The noise covariance matrix \(\boldsymbol{\Sigma}\) is here taken to be a general positive definite matrix, but may also be restricted to a diagonal matrix.
The goal of the model in \eqref{multioutGP} is to capture correlation of the log scores between experts.
This correlation is coming from two sources: in the noise distribution via \(\boldsymbol{\Sigma}\), and in the underlying GP prior for \(\v{f}(\v{z})\) with correlation generated by \(\m{C}\), which we now describe.

The model in \eqref{multioutGP} for all log scores in the training sample can be written
\begin{align}\label{multimodel}
    \underset{n\times K}{\boldsymbol{\ell}''} &=
        \underset{n\times K}{\m{F}} +
        \underset{n\times K}{\m{E}},
\end{align}
where \(\boldsymbol{\ell}''\) is the matrix consisting of the transformed log scores of \(K\) experts at \(n\) time points, \(\m{F}\) is determined by the multi-output GP, and \(\m{E}\) is a matrix of Gaussian noise terms. 

Reformulating the model in vectorized form, we can write the vector of errors as
\({\vectorize \m E = \boldsymbol{\Sigma} \otimes \m{I}_n}\). By letting \(\boldsymbol{\Sigma}\) be an arbitrary covariance matrix we allow correlation in the error terms across experts, but assume independence between observations.

Modeling the correlation in the underlying predictive ability is done via the GP prior.
A first attempt for a model for \(\m{F}\) is a matrix normal 
\begin{equation}\label{matrixnormal}
     \m{F} \sim \operatorname{\mathcal{MN}}(\v M, \v G,\v \Omega),
\end{equation}
i.e. that \(\vectorize \m{F}\) follows a multivariate normal with Kronecker-structured covariance matrix \(\m\Omega \otimes \m G\).
The matrix $\m G$ models the correlation between rows of $\m F$, which correspond to observations, and is therefore typically a covariance matrix generated from a kernel function, for example the squared exponential in \eqref{squared_exponential}.
The matrix $\m \Omega$ models the correlation between columns, in our case experts, and can most naturally be taken as a general positive definite covariance matrix. The signal variances in the squared exponential kernel, $\alpha$, would then be set to $1$ for each expert.

While the matrix normal model in \eqref{matrixnormal} has some intuitive appeal and is easy to work with from a computation perspective, it assumes that the columns of $\m F$ all follow the same multivariate normal distribution, which puts the restriction on our model that all experts have identical covariance function, i.e. the same covariance matrix $\m G$.
To be able to specify separate covariance functions for each expert we instead induce a prior for \(\m F\) by correlating a set of \(K\) underlying independent GPs, $\v h_1,\ldots,\v h_K$, each with its own kernel function \citep{teh_semiparametric_2005}
\begin{equation}\label{tilde_ver}
    \m F  = \m H \m C,
    \qquad
    \m H = \left[ \v h_1 \dots \v h_K \right],
    \qquad
    \v h_k \overset{\mathrm{indep}}{\sim} \N\left[\v 0, \operatorname G_k(\m Z, \m Z)\right],
\end{equation}
where \(\operatorname G_k(\m Z, \m Z^{\prime})\) is the covariance matrix for expert \(k\) generated from kernel function $g_k(\v z,\v z^\prime)$ and $\m C$ is a $K \times K$ matrix that generates the dependence between the $\v f_k$ for the experts.
We set the signal standard deviation to unity in all kernels, i.e. $\alpha_1=\cdots=\alpha_K=1$ in the squared exponential kernel in \eqref{squared_exponential} so that the scales of $\m F$ is generated by the $\m C$ matrix.
While the decision maker does not need to use the same pooling variables for each expert, we let the matrix of pooling variables \(\m Z\) contain the pooling variables for all expert to simplify notation.

The model in \eqref{tilde_ver} allows the decision maker to specify completely different covariance functions for the experts.
Importantly, this allows the length scales to differ between the expert in the case where an ARD kernel is used, but it is also possible to use, for example, a Matérn kernel \citep{rasmussen_gaussian_2006} for some of the experts and squared exponential for others.
It is also straightforward to force a subset of experts to share the exact same covariance function.

The covariance function for \(\m F\) in \eqref{tilde_ver} is a weighted sum of the covariance functions of the \(K\) experts
\begin{equation}\label{MV_covform}
    \operatorname{cov}(f_{k}(\v z_i), f_{l}(\v z_j)) = \sum_{s = 1}^K c_{sk}c_{sl} \mathrm{g}_s(\v z_i, \v z_j),
\end{equation}
where \(c_{kl}\) is element \((k,l)\) of the matrix \(\m C\).
To fully specify the multi-output \(\GP(\nicefrac{1}{3})\) model --- which we will denote by \(\operatorname{multi-GP}(\nicefrac{1}{3})\) --- we need to specify the covariance function of each expert, including priors for any unknown hyperparameters, as well as a prior for the matrix \(\m C\) that describes the between-expert covariances.

Note that while we choose to select a number of underlying processes equal to the number of experts, there is nothing that prevents us from creating our multi-output process by correlating an arbitrary number of latent GPs.
A common justification for using several models is that it gives a more holistic quantification of uncertainty by capturing the structural uncertainty of the parametric form of the models, rather than just the parametric uncertainty within models \citep{draper_assessment_1995}. 
From this perspective, including several models that share the majority of structural DNA can be problematic, for example in Bayesian model averaging \citep{hoeting_bayesian_1999}, where adding an effective "copy" of an already existing model while using a uniform prior will essentially double the posterior model weight of that model.
When modeling the joint predictive ability of a large pool of highly correlated experts it is possible that many of the elements of \(\m C\) will be close to zero, in which case restricting the number of underlying GPs could simplify the model and speed up calculations.

\subsubsection*{Inference for the multi-output \(\operatorname{GP}(\nicefrac{1}{3})\) model}

When modeling local predictive ability jointly, the end goal is the joint posterior of the ELPD for all experts at a new point \(\newobs{\mathbf{z}}\) in the pooling space, $\boldsymbol{\eta}(\v{z}) = (\eta_1(\newobs{\v{z}}),\ldots,\eta_K(\newobs{\v{z}}))^\top$, i.e. the mean of \(\boldsymbol{\ell}\) at \(\newobs{\mathbf{z}}\). Since \(\E(\boldsymbol{\ell}_{i})=\left(\E(\ell_{i1}), \dots, \E(\ell_{iK}) \right)^{\intercal}\) and \(\ell_{ik} = a_{ik} - \left(\ell_{ik}^{\prime\prime}\right)^3\), we can use the same formula for the third moments of Gaussians as in the univariate case and obtain the local predictive ability from model \eqref{multioutGP} pointwise as
\begin{equation}\label{eq:ELPDcuberoot_multi}
    \eta_k(\newobs{\v{z}}) = \newobs{a}_k - f_k^3(\newobs{\v{z}}) - 3f_k(\newobs{\v{z}})\boldsymbol{\Sigma}_{k, k},
\end{equation}
where \(\boldsymbol{\Sigma}_{k, k}\) denotes the \(k\)th diagonal element of the noise covariance matrix, and \(f_k(\newobs{\v{z}})\) the \(k\)th output of the GP at \(\newobs{\v{z}}\).

According to Equation \eqref{eq:ELPDcuberoot_multi} we obtain the posterior for $\boldsymbol{\eta}(\newobs{\v{z}})$ through the joint posterior ${p\left(\v{f}(\newobs{\v{z}}),\boldsymbol{\Sigma}, \boldsymbol{\Theta}, \m C \mid \boldsymbol{\ell}^{\prime\prime}\right)}$, where \(\boldsymbol{\ell}^{\prime\prime}\) is the $n \times K$ matrix of observed (transformed) log scores for all experts.
This posterior is expensive to sample from. However,
the cube root transformation makes it possible to extend the results from the univariate case to analytically integrate out $\m F$ in the training data , and obtain the marginal posterior of $\boldsymbol{\Sigma}$ and the GP hyperparameters ${\boldsymbol{\Theta} \text{ and } \m{C}}$ in closed form using a multivariate version of Eq.~2.8 in \citet{rasmussen_gaussian_2006}
\begin{equation}\label{eq:marglikegaussianGPreg_multi}
    p\big(\v \Sigma,\boldsymbol{\Theta}, \m C \mid \boldsymbol{\ell}^{\prime\prime}\big) = \int p\big(\m F,\boldsymbol{\Sigma},\boldsymbol{\Theta}, \m C \mid \boldsymbol{\ell}^{\prime\prime}\big)\,d\m F\;.
\end{equation} 
This lower-dimensional marginal posterior can be sampled with Hamiltonian Monte Carlo (HMC) using Stan \citep{stan_development_team_stan_2024, gabry_cmdstanr_2022} in a fraction of the time that it takes to sample the full joint posterior.

For each HMC sampled \((\boldsymbol{\Theta}, \m C, \boldsymbol{\Sigma})\) from \eqref{eq:marglikegaussianGPreg_multi} we can now easily sample from the distribution of \(\v{f}(\newobs{\v{z}})\) using \citep*{rasmussen_gaussian_2006}
\begin{equation}\label{eq:predictivefstar_multi}
    \v{f}(\newobs{\v{z}})
    \mid 
    \boldsymbol{\ell}^{\prime\prime}, \m{Z}, \newobs{\v{z}}, \boldsymbol{\Sigma}, \boldsymbol{\Theta} , \m C
    \sim 
    \N\left[
        \bar{\tilde{\v{f}}}, \operatorname{var}(\tilde{\v{f}}) 
    \right],
\end{equation}
where \(\m{Z}\) is the matrix of pooling variables in the training data, and
\begin{align*}
    \bar{\tilde{\v{f}}} &= \boldsymbol{\mu}(\newobs{\v z}) + 
    \operatorname{G}(\newobs{\v{z}}, \m{Z})
    \left[
        \operatorname{G}(\m{Z}, \m{Z}) + \boldsymbol{\Sigma} \otimes \m I_n
    \right]^{-1} (\vectorize\boldsymbol{\ell}^{\prime\prime} - \boldsymbol{\mu}),\\
    \var(\tilde{\v{f}}) &=
    \operatorname{G}(\v{z}, \newobs{\v{z}}) -
    \operatorname{G}(\newobs{\v{z}}, \m{Z})
    \left[
        \operatorname{G}(\m{Z}, \m{Z}) + \boldsymbol{\Sigma} \otimes \m I_n
    \right]^{-1}
    \operatorname{G}(\m{Z}, \newobs{\v{z}}),
\end{align*}
where, for example, \(\operatorname G(\m{Z}, \m{Z})\) is the \(Kn\times Kn\) matrix with covariances of the form \(\operatorname{cov}(f_k(\v{z}_i),f_l(\v{z}_j))\) for experts \(k\) and \(l\) and observations \(i\) and \(j\) in the training data.
The posterior draws of \(\boldsymbol{\Sigma}\) from \eqref{eq:marglikegaussianGPreg_multi} and \(\v{f}(\newobs{\v{z}})\) from \eqref{eq:predictivefstar_multi} can finally be inserted into \eqref{eq:ELPDcuberoot_multi} to obtain draws from the posterior of the local ELPD \(\boldsymbol{\eta}(\newobs{\mathbf{z}})\).
The method is summarized in Algorithm \ref{alg:GPcuberootjoint}. 

\begin{algorithm}
    \caption{Posterior sampling for ELPD at $\tilde{\mathbf{z}}$ in the Multi-\(\GP\left(\nicefrac{1}{3}\right)\) model}\label{alg:GPcuberootjoint}
    \KwIn{Number of posterior draws $M$, a vector of pooling variables $\tilde{\mathbf{z}}$ and $\tilde a_{k}=-(1/2)\log(2\pi \tilde \sigma_k^2)$ for expert $k$.
    }
    \vspace{2mm}
  
  Generate \(M\) draws \(\{\boldsymbol{\Sigma}^{(j)}, \boldsymbol{\Theta}^{(j)}, \m{C}^{(j)}\}_{j=1}^M\) from \(p\big(\boldsymbol{\Sigma},\boldsymbol{\Theta}, \m{C} \mid \boldsymbol{\ell}^{\prime\prime}\big)\) in \eqref{eq:marglikegaussianGPreg_multi} using HMC.
  \vspace{1mm} \\
  \For {\(j = 1,\dots, M\)}{
  Sample the posterior $\v{f}^{(j)}(\newobs{\v{z}}) \mid 
    \boldsymbol{\ell}^{\prime\prime}, \m{Z}, \newobs{\v{z}}, \boldsymbol{\Sigma}^{(j)}, \boldsymbol{\Theta}^{(j)}, \m{C}^{(j)}$ in \eqref{eq:predictivefstar_multi}\\
    Compute joint local predictive ability $\boldsymbol{\eta}^{(j)}(\newobs{\v{z}})$ from \eqref{eq:ELPDcuberoot_multi}
  \\ \vspace{1mm}
  }
  \vspace{1mm}
 \KwOut{Sample from the joint local predictive ability posterior: $\boldsymbol{\eta}^{(1)}(\newobs{\v{z}}), \boldsymbol{\eta}^{(2)}(\newobs{\v{z}}), \ldots, \boldsymbol{\eta}^{(M)}(\newobs{\v{z}})\;.$}
\end{algorithm}

\section{Predicting bike-sharing utilization rates in Washington D.C.}\label{sec_empirical}

The last decade has seen an explosion in companies offering short-term rentals of final mile transportation, such as bikes or electric scooters.
Key for the success of such a business is to be able to correctly predict demand.
In this section we use single- and multi-output GP models to estimate the joint one-step-ahead local predictive ability of three experts predicting bike rentals based on the data in \citet*{fanaee-t_event_2014}.
We then propose novel methods for how to use the estimates of local predictive ability to form aggregate predictions, and evaluate the single- and multi-output versions with regards to out-of-sample performance.

\subsection{Data and models}

The dataset in \citet*{fanaee-t_event_2014} includes the daily number of bike rentals over the two year period from January 1, 2011, to December 31, 2012.
The dataset further contains several covariates relating to the weather, as well as indicators for official US holidays and an indicator for if the day in question is a workday.

We use the same three experts as in \citet{oelrich_local_2024} to generate predictions: a Bayesian regression model (BREG), a Bayesian additive regression tree model (BART), and a Bayesian linear regression model with stochastic volatility (SVBVAR).
These experts use several weather-related variables, an indicator for season, the number of rentals the previous day, and indicators for workday and holiday as covariates.
For further discussion of the data see \citet*{oelrich_local_2024}.

As pooling variables the decision maker uses humidity, wind speed, and temperature from the dataset in \citet*{fanaee-t_event_2014}.
The decision maker also decides to use a variable she calls \emph{family holiday} which is not included in the dataset, or in the estimation of any of the experts, which takes the value \(1\) at Thanksgiving and Christmas (Eve and Day) \citep*{oelrich_local_2024}.
This variable is included to represent the decision makers belief that there are certain stay-at-home holidays where the demand for rental bikes is almost non-existent.
As neither of the experts have access to this variable, she expects their predictive ability to vary significantly with regards to this variable. 

\subsection{Estimating predictive ability}

To estimate local predictive ability, we use the \(\GP\left(\nicefrac{1}{3}\right)\) method, as the number of data points makes the \(\GP(\chi^2_1)
\) method inconveniently slow. More specifically, we use a GP with a squared exponential kernel with automatic relevance determination (ARD) so that each pooling variable is associated with its own length scale hyperparameter \citep{rasmussen_gaussian_2006}. Following the Gaussian process regression example in the Stan User's Guide, we give the length scales independent \(\operatorname{Inverse-Gamma}(5,5)\) priors, and the noise and signal variance are given \(\N^+(0, 1)\) priors \citep{stan_development_team_stan_2024}. 

\begin{figure}
    \centering
    \includegraphics[width=0.9\textwidth]{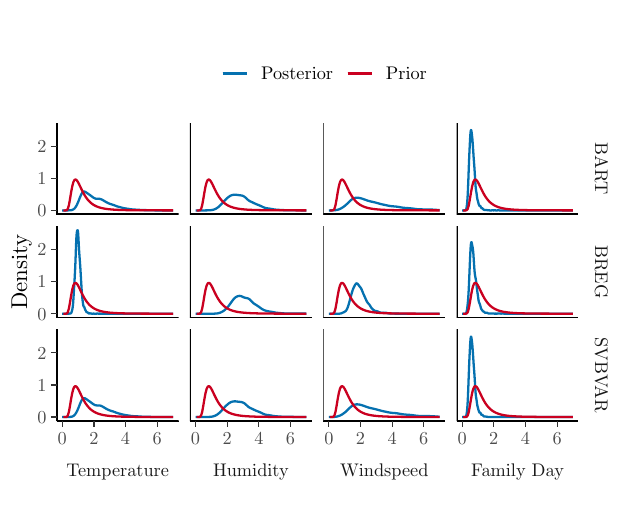}
    \caption{Marginal posterior and prior distributions of the length scales for all three experts at the final time point.}
    \label{fig_length_scales}
\end{figure}

Figure \ref{fig_length_scales} shows the prior and posterior distribution of the length scales for all three experts at the final time point. The length scale tells us how quickly the correlation between two points decreases, so smaller length scales are indicative of greater locality, in the sense of local predictive ability varying more rapidly over the pooling space. \emph{Temperature} and in particular \emph{Family Day} show the strongest evidence of locality with a posterior that is heavily concentrated on smaller length scales. There is also some variation between the experts, with the Bayesian regression model showing more evidence of local behavior.

\begin{figure}
    \centering
    \includegraphics{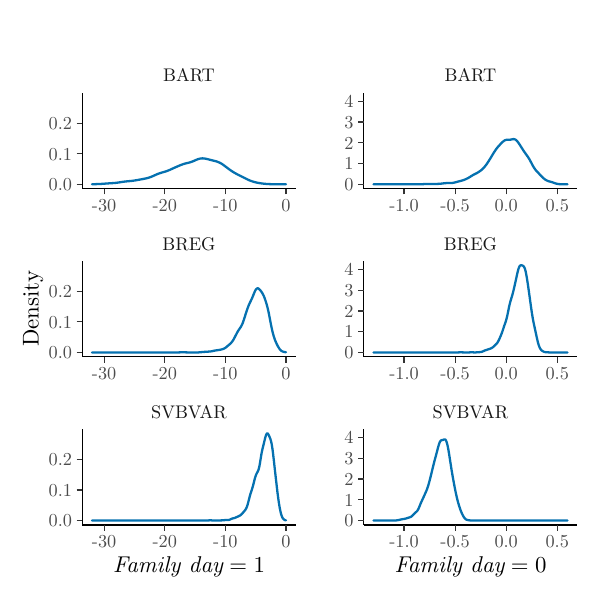}
    \caption{The left side shows the posterior distribution of local predictive ability for Christmas Day \(2012\). The right side shows the same distribution but with \emph{family day} set to \(0\).}
    \label{etaplot}
\end{figure}

To illustrate how local predictive ability changes over the pooling space, Fig. \ref{etaplot} shows the actual posterior of local predictive ability for Christmas Day \(2012\) together with what the posterior would have looked like if we keep all pooling variables constant except the value of \emph{family day}. The BART model would, if this had not been a family day, been expected to perform roughly as well as the Bayesian regression model, with the stochastic volatility expected to have the poorest performance. However, for \(\textit{family day}=1\), which is what was observed, the model expects the stochastic volatility model to perform well and the BART to have the worst local predictive ability by a wide margin. 

\begin{figure}
    \centering
    \makebox[\textwidth][c]{\includegraphics[width=0.9\textwidth]{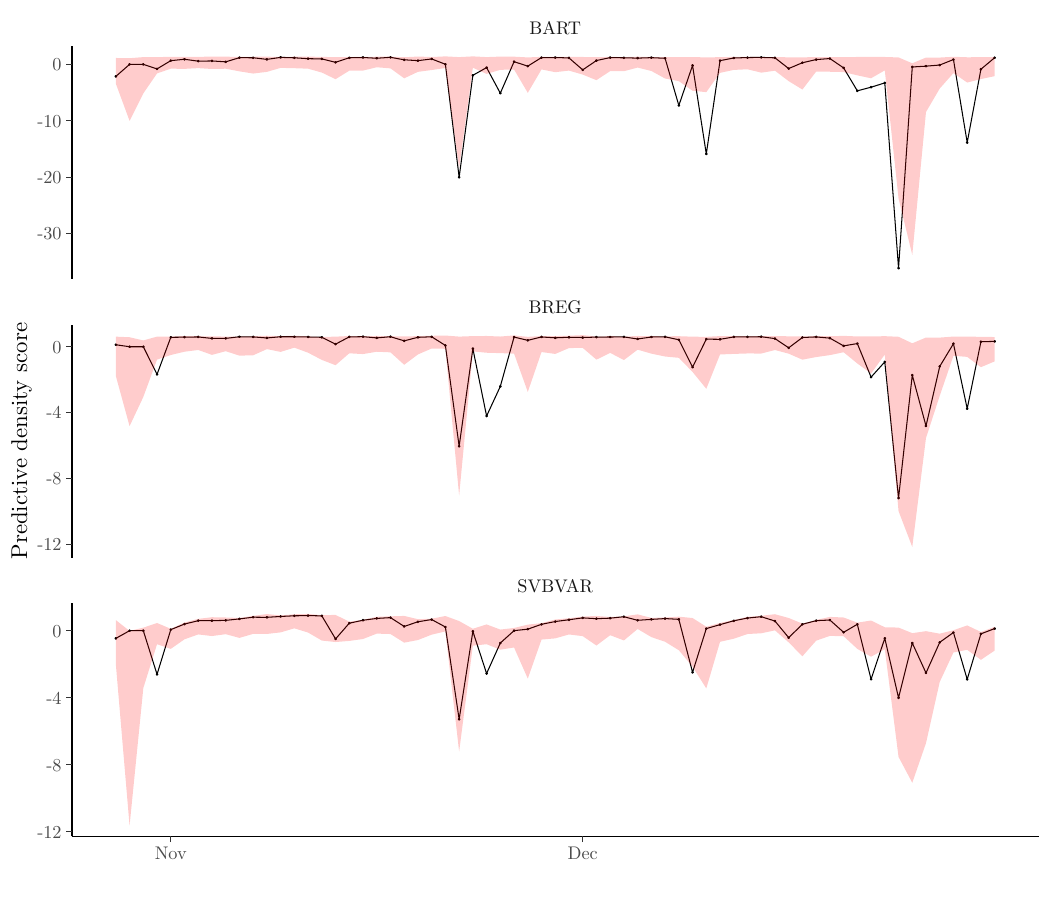}}
    \caption{Estimates of local predictive ability for a subset of the bike sharing data. The black line shows the observed log scores and the red ribbon shows a \(95 \%\) HPD interval for the posterior predictive distribution of log scores.}
    \label{lpa_main}
\end{figure}

Figure \ref{lpa_main} shows a \(95 \%\) highest posterior density (HPD) interval for the predictive distribution of a subset of the log scores. For all experts,  the majority of observed log scores land within the HPD. However, around twice as many as would be expected fall outside, with BART doing the worst and SVBVAR the best. This may indicate that the decision maker could be missing an important pooling variable.

\begin{table}[ht]
\centering
\caption{Mean and median log predictive score of one-step-ahead predictions of log scores. See \eqref{eq:benchmarkmodels} for a description of the models.}
\begin{tabular}{llrr}
  \hline
 Expert & Method & Median & Mean \\ 
  \hline
 BART & $\ell^{'}$ random walk & -1.80 & -3.14 \\ 
    & $\ell^{'}$ cumulative mean & -1.83 & -2.62 \\ 
    & $\ell^{''}$ random walk & -0.20 & -0.58 \\ 
   & $\ell^{''}$ cumulative mean & -0.07 & $\mathbf{-0.42}$ \\ 
    & $\GP(\nicefrac{1}{3})$ & $\mathbf{0.02}$ & -2.10 \\ \\
  BREG & $\ell^{'}$ random walk & -1.08 & -1.80 \\ 
   & $\ell^{'}$ cumulative mean & -1.11 & -1.54 \\ 
   & $\ell^{''}$ random walk & 0.23 & -0.03 \\ 
   & $\ell^{''}$ cumulative mean & 0.44 & $\mathbf{0.13}$ \\ 
   & $\GP(\nicefrac{1}{3})$ & $\mathbf{0.58}$ & -3.67 \\ \\
  SVBVAR & $\ell^{'}$ random walk & -1.38 & -1.81 \\ 
   & $\ell^{'}$ cumulative mean & -1.40 & -1.61 \\
   & $\ell^{''}$ random walk & 0.19 & -0.14 \\ 
   & $\ell^{''}$ cumulative mean & 0.58 & $\mathbf{0.20}$ \\ 
   & $\GP(\nicefrac{1}{3})$ & $\mathbf{0.68}$ & -2.27 \\ 
   \hline
\end{tabular}
\label{tbl_lpa_eval}
\end{table}

Table \ref{tbl_lpa_eval} evaluates the quality of the one-step-ahead predictions of local predictive ability from the \(\GP(\nicefrac{1}{3})\) model and compares it with four benchmark models: 
\begin{align}\label{eq:benchmarkmodels}
    \ell' \text{ random walk} &:\hspace{0.5cm} \ell'_{ik} \mid \ell'_{i-1,k} \sim \N(\ell'_{i-1,k},\sigma^2) \nonumber \\
    \ell' \text{ global mean} &:\hspace{0.5cm} \ell'_{ik} \sim \N(\mu,\sigma^2) \nonumber \\
    \ell'' \text{ random walk} &:\hspace{0.5cm} \ell''_{ik} \mid \ell''_{i-1,k} \sim \N(\ell''_{i-1,k},\sigma^2) \nonumber \\
    \ell'' \text{ global mean} &:\hspace{0.5cm} \ell''_{ik} \sim \N(\mu,\sigma^2),
\end{align}
where all models are finally transformed to predict the original log scores $\ell_{ik}$. Table \ref{tbl_lpa_eval} shows that the two reference methods based on the linearly transformed log scores $\ell'$ perform badly across the board, giving additional support to the cube root transformation. The global model based on power-transformed log scores is the best performer in terms of the mean, but \(\GP(\nicefrac{1}{3})\) outperforms it when it comes to the median. This discrepancy between mean and median is due to a few extreme outliers, the worst being a log score of roughly \(-700\). 

\begin{figure}
    \centering
    \includegraphics[width=0.8\textwidth]{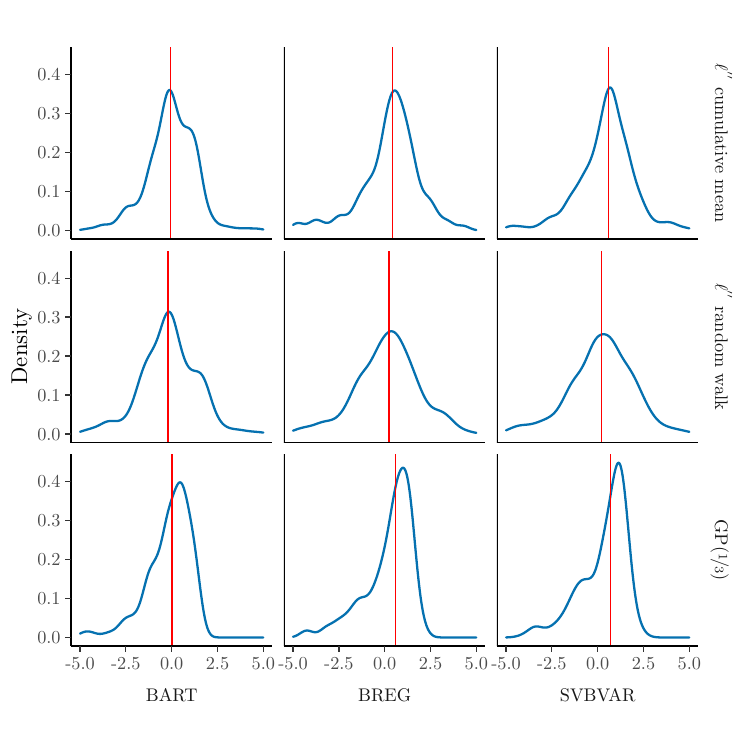}
    \caption{Kernel density estimates of one-step-ahead log density evaluations of the predicted log scores, $\log p(\ell_{i+1} \mid \ell_{1:i})$, in the test data for the bike rental data. The rows in the figure corresponds to predictions from the \(\GP(\nicefrac{1}{3})\) model and two global reference methods, respectively. The median of each distribution is marked out with a red vertical line. }
    \label{logscore_logscore_fig}
\end{figure}

Figure \ref{logscore_logscore_fig} shows how well the power-transformed benchmark models and the \(\GP(\nicefrac{1}{3})\) model predicts the observed log scores for each expert. The figure shows (kernel density estimates of) the log density of all observed log scores, ${\log p(\ell_{i+1} \mid \ell_{1:i})}$, in the test data, with respect to each local ELPD model. The median of each distribution is marked out with a red vertical line. We see that the \(\GP(\nicefrac{1}{3})\) model performs better on a majority of the observations, but has a longer left tail caused by a very low predictive density for some observed log scores.


\begin{figure}
    \centering
    \includegraphics {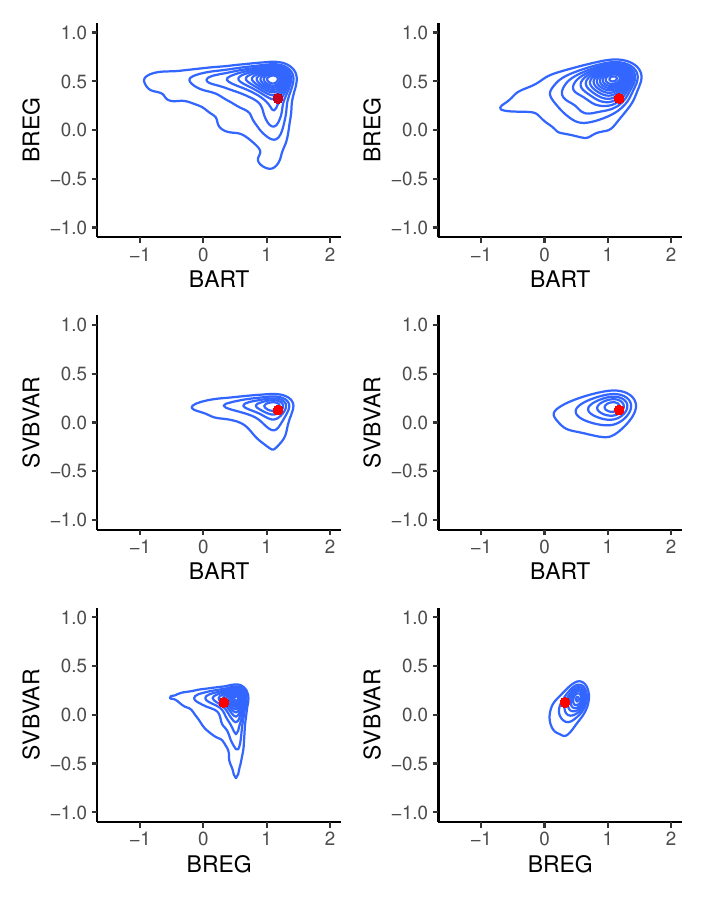}
    \caption{Joint predictive distributions of log scores for the last data point in the bike-sharing data. The graphs show kernel density estimates for univariate (left side) and multivariate (right side) Gaussian processes. The observed log scores are marked with a red dot.}
    \label{fig:ls_preddist}
\end{figure}

To estimate the joint predictive ability of all three experts we use the \(\operatorname{multi-GP}(\nicefrac{1}{3})\) model, also with a squared exponential automatic relevance determination (ARD) kernel \citep*{rasmussen_gaussian_2006}.
To allow for the possibility that some pooling variables are unimportant to some experts we use a fairly flat prior for the length scales, \(\operatorname{Cauchy}(0, 5)\), restricted to the interval \((0, 100)\).
For the \(\m C\) matrix we use a \(\mathrm{LKJ}(\eta = 3)\) prior \citep{lewandowski_generating_2009}, and zero-truncated \(\N^+(0, 1)\) priors for the signal and error term variances.

Figure \ref{fig:ls_preddist} displays the bivariate joint predictive distribution of each expert pair at the last point in the data set, with the first column showing the joint predictive distributions from single output GPs --- implicitly assuming independence --- and the second column showing the joint distributions based on the multi-output model.
The multi-output model captures positive correlations between all experts, especially between the Bayesian regression model and the stochastic volatility model.
The smaller correlations with BART is is not surprising, since the BART model is a quite different model with possibly highly nonlinear mean function.
Note also that the predictive variances for the multi-output distributions are \(10\)--\(20\%\) lower at this time point.

\subsection{Forecast combination based on estimates predictive ability}

There are many reasons to model joint predictive ability.
For example, it can be done purely for inference purposes, such as when we are interested in examining how the quality of predictions for different expert co-vary over a pooling space, or as a part of model evaluation.
In this section, we focus on how to create a linear prediction pool \citep*{geweke_optimal_2011} based on these estimates.

A linear prediction pool can be made local by letting its weights depend on \(\v z\), so that the predictive density at a new point \(\newobs{\v z}\) in the pooling space is given by
\begin{equation}
    p\left(\tilde y \mid \newobs{\v z}\right) = \sum_{k=1}^{K} w_k(\newobs{\v z}) p_k(\tilde y), \quad \sum_{k=1}^K w_k(\newobs{\v z}) = 1, \quad w_k(\newobs{\v z}) \ge 0,
\end{equation}
where \(w_k(\newobs{\v z})\) is the weight of expert \(k\) at \(\newobs{\mathbf{z}}\). To bring the predictive ability of the expert into the pooling we let the weight \(w_k(\newobs{\v z})\) be a function of the posterior probability that expert \(k\) has the highest predictive ability at \(\newobs{\v z}\).
We denote this probability by \(\psi_k(\newobs{\v z}) = p\big(\eta_k(\newobs{\v z}) > \eta_{j \ne k}(\newobs{\v z})\big)\), and refer to a forecast combinations based directly on \( \psi_k(\newobs{\v z})\) as \(\GP(\nicefrac{1}{3})\) \emph{natural} in tables and figures.

Often, and especially when the amount of data is small, these weights will tend to not discriminate strongly between experts and be outperformed by more aggressive weighing schemes such as the optimal linear pools of \citet{geweke_optimal_2011}. We therefore propose scaling \(p\big(\eta_k(\tilde{\mathbf{z}}) > \eta_{j \ne k}(\tilde{\mathbf{z}})\)\big) by a constant \(c\) and then feeding it through a softmax transformation
\begin{equation}\label{softmax_with_discrim}
    w_k(\newobs{\v z}) =
    \frac{
        \exp
        \left( 
            c \cdot \psi_k(\newobs{\v z}) 
        \right)
    }{
        \sum_{j=1}^K 
        \exp
        \left( 
            c \cdot \psi_j(\newobs{\v z})
        \right)
    }.
\end{equation}
The constant \(c\), which we will refer to as the discrimination factor, allows the decision maker to tune how aggressively to discriminate between the experts. Setting \(c = 0\) will result in equal weights, and setting \(c = \infty\) will assign weight one to the expert with the highest probability of having greater predictive ability than all other experts. Typically a value of \(c\) between these extremes will yield the best results.

When the decision maker has no strong prior opinion of \(c\), it can be dynamically selected at each time point by selecting the value that optimizes the historical predictive ability of the pool. We refer to this method as \(\GP(\nicefrac{1}{3})\) \emph{dynamic} in tables and figures. As a special case of interest, we will include \(c = \infty\) under the name \(\GP(\nicefrac{1}{3})\) \emph{model selection}, as it assigns weight \(0\) to all but one of the experts.

\begin{figure}
    \centering
    \includegraphics[scale=0.7]{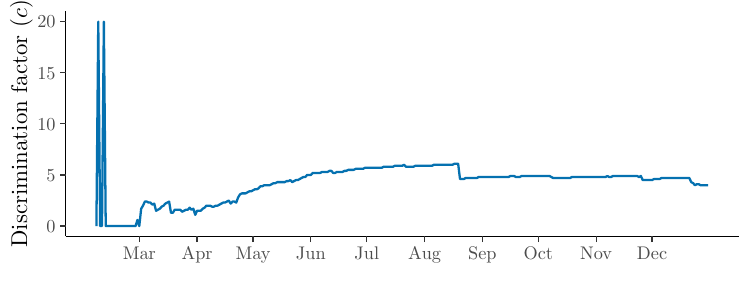}
    \caption{Change in optimal discrimination factor over time, single output.}
    \label{fig:opt_c}
\end{figure}

Figure \ref{fig:opt_c} shows how the optimal discrimination changes over time for the bike share data for the single-output method. At each time point, the historical performance of the pool for \(0 \le c \le 20\) is evaluated, and the next prediction is made with the value of $c$ corresponding to the historically most accurate pool. The optimal discrimination $c$ is volatile when little historical data is available, but then settles down around $c=5$ as a compromise between the extremes of equal weights and assigning all weight to the expert with highest probability of having the best predictive ability at $\tilde{\mathbf{z}}$.

\begin{figure}
    \centering
    \includegraphics[scale=0.7]{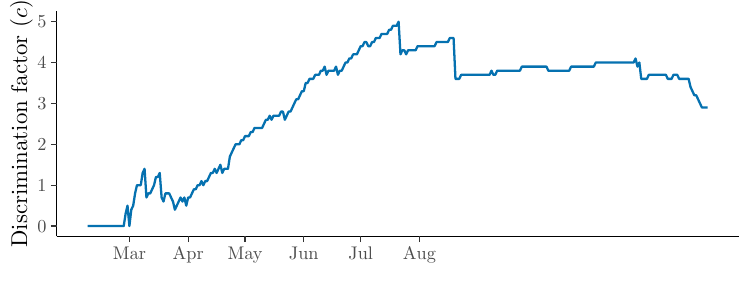}
    \caption{Change in optimal discrimination factor over time, multi-output.}
    \label{fig:opt_c_multi}
\end{figure}

The optimal discrimination factor for the multi-output model, illustrated in Fig. \ref{fig:opt_c_multi}, settles around a lower value than in the single output version. This is due to the single-output methods failure to capture the strong positive correlation between the predictive ability of the experts, which naturally leads to a lower degree of discrimination. This is discussed further in Sect. \ref{corr_exp_disc}.

\begin{figure}
    \centering
    \includegraphics[width=1\textwidth]{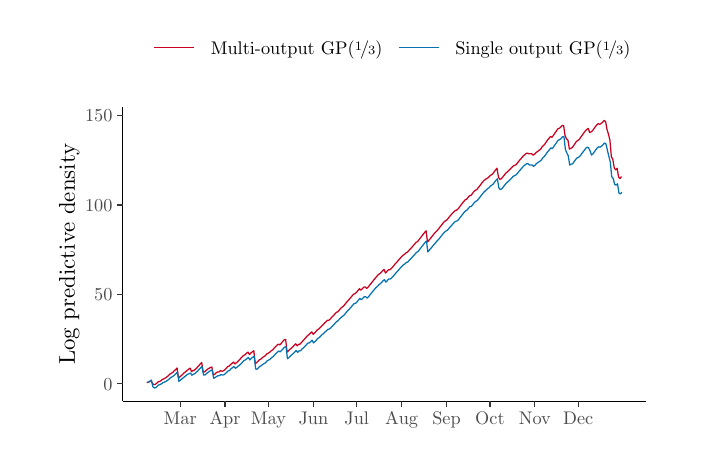}
    \caption{Cumulative log predictive scores based on the natural aggregation scheme for single and multi-output models.}
    \label{fig:aggpreds_thompson_cumu}
\end{figure}

\begin{figure}
    \centering
    \includegraphics[width=1\textwidth]{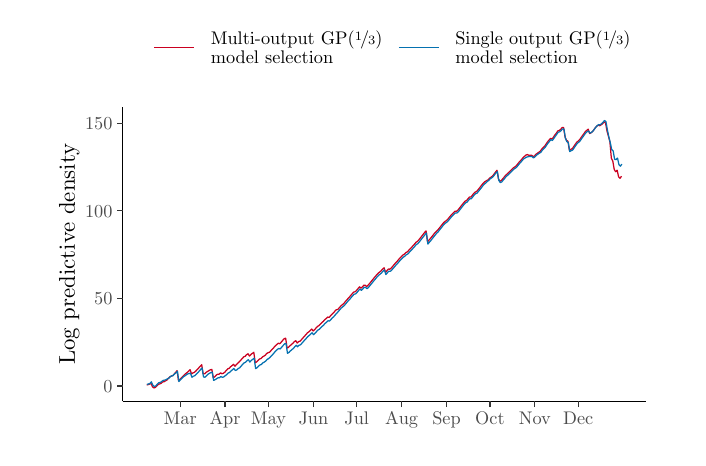}
    \caption{Cumulative log predictive scores based on the model selection aggregation scheme for single and multi-output models.}
    \label{fig:aggpreds_selbest_cumu}
\end{figure}

\begin{figure}
    \centering
    \includegraphics[width=1\textwidth]{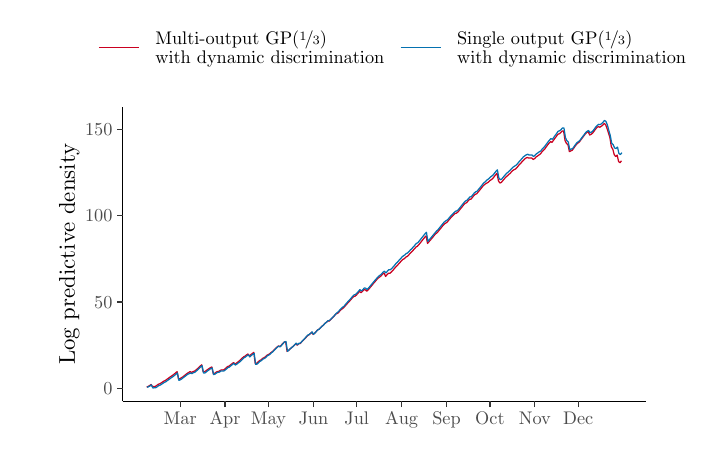}
    \caption{Cumulative log predictive scores based on the dynamic aggregation scheme for single and multi-output models.}
    \label{fig:aggpreds_dynamic_cumu}
\end{figure}

All pooling methods described above will work regardless of whether the estimates of local predictive ability have been obtained with multiple single output GPs or a single multi-output GP, but taking account of the dependence will change the weights of the experts, sometimes drastically. 
This can be seen in Figs. \ref{fig:aggpreds_thompson_cumu}--\ref{fig:aggpreds_dynamic_cumu}, which compare the single- and multi-output approaches for the three aggregation methods discussed above.
The multi-output model clearly outperforms its single output counterparts when using the natural weighting scheme. When using model selection or dynamic discrimination, however, the simpler single-output GP method performs slightly better, in both cases almost entirely due to a few observations at the end of the dataset.

\subsection{Comparison with reference methods}

The sum of the log predictive scores over time for each method, together with the performance of the individual experts as well as a selection of reference methods, are shown in Table \ref{logscore_sums}. The reference methods included are: the caliper method used in \citet{oelrich_local_2024}, which is a nonparametric method that averages log scores of nearby observations in $\mathbf{z}$-space; the global optimal pool of \citet{geweke_optimal_2011} which selects aggregation weights that optimize the historical performance of the pool in terms of log scores; the local optimal pool in \citet{oelrich_local_2024}, which is an extension of the optimal pool of \citet*{geweke_optimal_2011} where the optimal weights are determined on a subset of the data in the pooling space closest to the current point. For details about these reference methods, see \citet{oelrich_local_2024}.

While the natural multi-output GP outperforms its single-output counterpart considerably, this is reversed in the dynamic version. The single-output dynamic \(\GP(\nicefrac{1}{3})\) performs the best out of the methods based on local predictive ability, but is soundly beaten by the local optimal pool of \citet{oelrich_local_2024}.

\begin{table}
\centering
\caption{Sum of log predictive density scores over the whole test period.}
\begin{tabular}{lr}
    \toprule
   Method & Sum of log scores \\ 
   \midrule
   SVBVAR & 45.1 \\ 
   BREG & 28.8 \\ 
   BART & 43.6 \\ 
   Equal weights & 98.7 \\
   Caliper & 133.7 \\ 
   Global opt. & 130.2 \\ 
   Local opt. & 145.3 \\
   \midrule
   $\GP(\nicefrac{1}{3})$ natural & 107.1 \\ 
   multi-$\GP(\nicefrac{1}{3})$ natural & 115.9 \\ 
   $\GP(\nicefrac{1}{3})$ model sel. & 126.6 \\ 
   multi-$\GP(\nicefrac{1}{3})$ model sel. & 119.7 \\
   $\GP(\nicefrac{1}{3})$ dynamic & 136.4 \\
   multi-$\GP(\nicefrac{1}{3})$ dynamic & 131.7 \\
     \bottomrule
  \end{tabular}
  \label{logscore_sums}
\end{table}

\begin{figure}[htbp]
    \centering
    \makebox[\textwidth][c]{\includegraphics[width=1.0\textwidth]{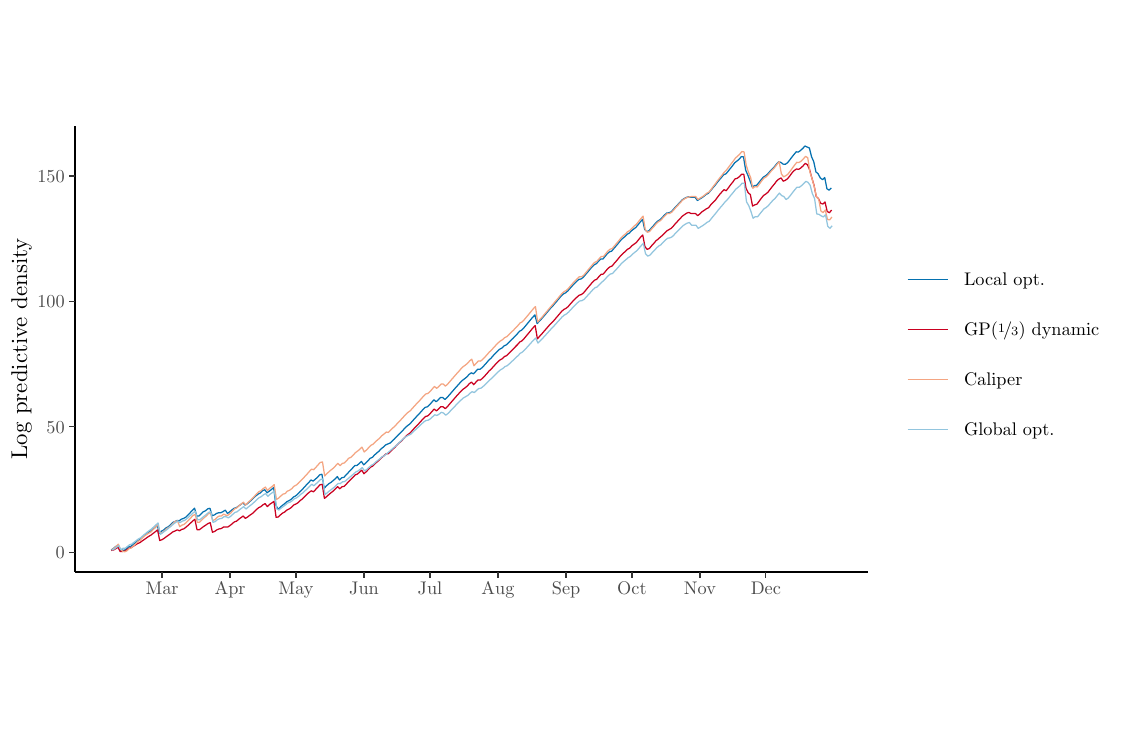}}
    \caption{Cumulative log predictive density for the four best methods. The legend is sorted in descending order of total log predictive score.}
    \label{aggpreds_ref}
\end{figure}

Figure \ref{aggpreds_ref} compares the best performing of the \(\GP(\nicefrac{1}{3})\) pooling schemes with the reference methods. The \(\GP(\nicefrac{1}{3})\) initially struggles when there is little training data, but then quickly improves over time and at the end of the sample it only outperformed by the local optimal pool. 

\subsection{Correlated experts and degree of discrimination}\label{corr_exp_disc}

\begin{figure}
    \centering
    \includegraphics[width=12cm]{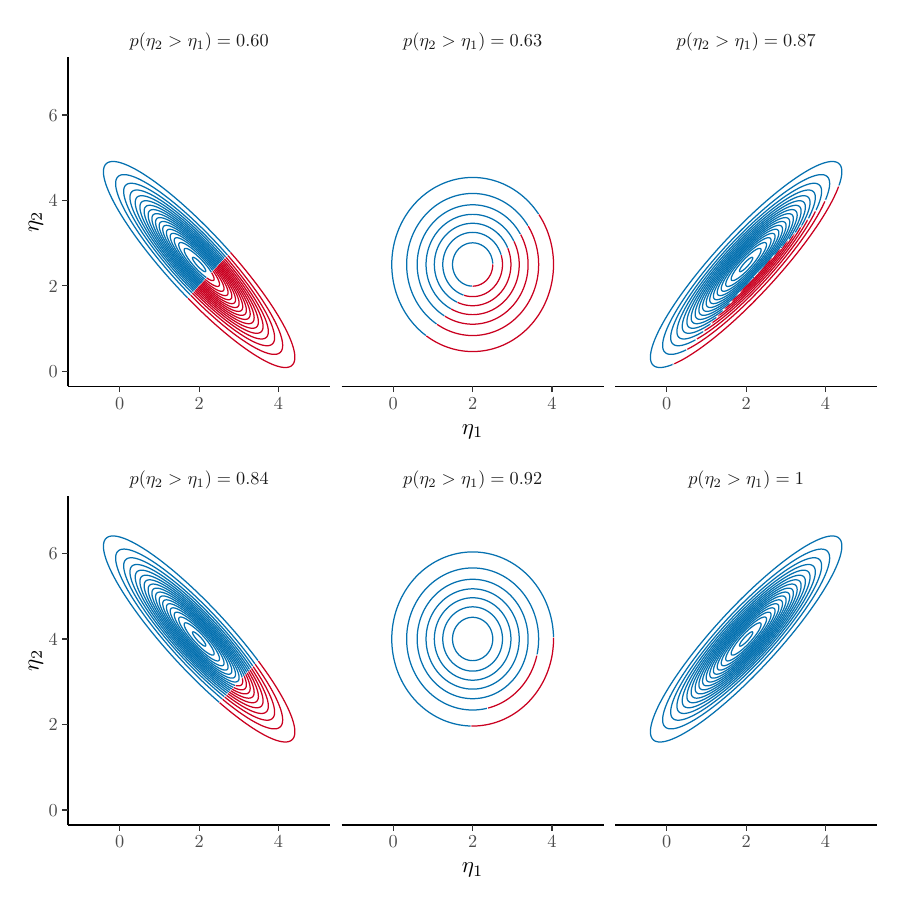}
    \caption{Gaussian contour plot illustrating the effect of correlation in predictive ability on the posterior probability that Expert 2 has greatest predictive ability at $\newobs{\v z}$. In the top row, the marginal difference in posterior mean of predictive ability between the two experts is small ($\E(\eta_2(\newobs{\v z}))-\E (\eta_1(\newobs{\v z})) =0.5$), and in the bottom row the difference is substantial ($\E(\eta_2(\newobs{\v z}))-\E(\eta_1(\newobs{\v z})) =2$). The blue part of each contour plot marks the parts of the joint density where \(\eta_2(\newobs{\v z}) > \eta_{1}(\newobs{\v z})\).}
    \label{fig:correlation_effectc}
\end{figure}

The multi-output model owes its superior performance when using natural weighting to a larger degree of discrimination, as the weak discrimination of the single-output method is in large part caused by a failure to capture correlation between the experts. A positive correlation that is unaccounted for will lead to weaker discrimination, and failure to take into account a negative correlation will lead to discrimination that is too strong. Typically, experts will be positively correlated and the degree of discrimination will be underestimated. This effect is illustrated in Fig. \ref{fig:correlation_effectc}.

Figure \ref{fig:correlation_effectc} displays contour plots of the posteriors for $\eta_1$ and $\eta_2$ for two experts in two setups.
In the top row, the difference in posterior mean of predictive ability between the two experts is small (${\E(\eta_2)-\E (\eta_1) =0.5}$), and in the bottom row the difference is more substantial ($\E(\eta_2)-\E(\eta_1) =2$).
In both setups we fix the marginal posterior variances of the two experts at \(1\), and explore what happens when we let the correlation go from strong and negative \((\rho = -0.9)\) to non-existent \((\rho = 0)\) to strong and positive \((\rho = 0.9)\).
As the correlation between the experts approaches one, the posterior probability of having the highest predictive ability will concentrate on the better model, no matter how small the difference in posterior means.
Compared to uncorrelated experts, negative correlation increase the posterior probability that the expert with lower posterior mean has the higher predictive ability, something which becomes clearer as the magnitude of the difference in mean increases.

The effect illustrated in Fig. \ref{fig:correlation_effectc} explains why the multi-output GP outperforms the single output version  when using natural weights. By capturing the correlation between experts, the multi-output GP leads to natural weights that discriminate more strongly between the experts.

\section{Conclusions and further research}\label{sec_conclusions}

We have introduced a method for estimating the predictive ability \(\eta(\mathbf{z})\) of a model or expert, in terms of ELPD, as it varies over a space of pooling variables \(\mathbf{z}\). We call this conditional predictive ability \emph{local predictive ability}, and propose using a Gaussian process to model it. 

Gaussian processes have the drawback of being computationally heavy, and so we suggest the use of a power transformation of log scores which allows us to use a much faster marginalized GP to generate draws from the posterior of \(\eta(\mathbf{z})\). Our proposed method is shown to work well for predicting bike share data \citep*{fanaee-t_event_2014}, where it outperforms all compared methods except the local version of the linear prediction pool of \citet*{geweke_optimal_2011} proposed in \citet{oelrich_local_2024}.

Our method can be used to both model the marginal distribution of the predictive ability \(\eta_k(\mathbf{z})\) for each expert, and to model the predictive ability of the experts jointly using a multi-output GP that makes it possible to capture correlation between the predictive abilities of the experts as well as correlation in the noise.

While the inference and prediction from the marginalized version is much faster compared to sampling also the latent variables in the GP($\nicefrac{1}{3}$) model, the step from running several single output GPs side by side to running a single multi-output GP increases the computational burden significantly, making large sample sizes time consuming.
A topic for further study is therefore methods for speeding up the calculation using sparse linear algebra, methods for large-scale homoscedastic Gaussian GPs \citep{liu_when_2020}, or using alternative Bayesian inference approaches than Monte Carlo sampling, such as variational inference based on optimization \citep{blei_variational_2017}.

Another potential avenue of future research is using alternative methods to pool predictions based on local predictive ability. In the empirical application we use three different methods, all of which have pooling weights based on \(\psi_k\), the posterior probability that a particular expert has greater predictive ability than all other experts. Alternatively, one can form linear combinations directly on the $\eta_1(\v z),\ldots,\eta_K(\v z)$ to obtain a posterior for the pooled local ability $\sum_{k=1}^K w_k \eta_k(\v z)$. 
Given a set of criteria for this distribution, such as maximizing the mean while keeping the variance below a certain threshold, a set of local weights can be obtained through numerical optimization. 

Allowing for time-varying weights has been shown to be beneficial in macroeconomic forecasting \citep{del_negro_dynamic_2014, billio_time-varying_2013, mcalinn_multivariate_2020}. While we have not explicitly discussed modeling the persistence of local predictive ability over time, it is straight forward to include some function of time in our vector of pooling variables via an appropriate kernel function.

\clearpage

\section*{Appendix}
\addcontentsline{toc}{section}{Appendix}
In this appendix we investigate the properties of our methods on simulated data. We are particularly interested in investigating how the  computationally faster $\GP(\nicefrac{1}{3})$ model compares to the computationally costly $\GP(\chi_1^2)$ model. 

We generate data from the model
\begin{equation}\label{simulation_dgp}
    y_i = x_{1i} + x_{2i} + \epsilon_i, \qquad \epsilon_i \overset{\mathrm{iid}}{\sim} \N(0, 1),
\end{equation}
where the covariates \(x_1\) and \(x_2\) are generated as i.i.d. standard normal variates. As our expert we use a Bayesian regression model that is misspecified in that one of the covariates is missing, i.e.
\begin{equation}\label{simulation_expert}
    y_i = \alpha + \beta x_{1i} + \varepsilon_i, \text{ where } \varepsilon_i \overset{\mathrm{iid}}{\sim}\N(0,\sigma_{\varepsilon})\;.
\end{equation}
To make the expert model as simple as possible for this illustration, we assume that it has been trained with so much data that the posterior of all coefficients will have variances close to zero and posterior means equal to \(0\) for \(\alpha\), \(1\) for \(\beta\), and \(2\) for \(\sigma_{i}^2\) (since \(\varepsilon_i\) absorbs both \(\epsilon_i\) in the DGP and the variation from the missing \(x_{2i}\)).

We further assume that our decision maker believes that the predictive ability of the expert varies over the variable \(x_2\), but does not know the exact relationship between the ELDP \(\eta\) and \(x_2\). We will examine how well the decision maker will be able to estimate the true predictive ability, as summarized by the posterior distribution of \(\eta\), using the two models described in Sect. \ref{sec_framework}.

The true distribution of the log scores in this setup is given by
\begin{equation*}
    \ell_i \overset{d}{=} -\frac{1}{2}\log 2\pi\sigma_i^2 - \frac{1}{2 \sigma_i^2} (y_i - \mu_i)^2,
\end{equation*}
with each \(y_i \mid x_{1i}, x_{2i} \overset{\mathrm{indep.}}{\sim} \N(x_{1i} + x_{2i},1)\). 
For our expert \(\mu_i = x_{1i}\) and \(\sigma_i^2 = 2\), so this simplifies to
\begin{equation*}
    \ell_i \overset{d}{=} -\frac{1}{2}\log 4\pi - \frac{1}{4}(x_{2i} + \epsilon_i)^2\;.
\end{equation*}
As \(x_{2i}+\epsilon \sim \N(x_{2i}, 1)\), it follows that  \((x_{2i}+\epsilon)^2 \sim \chi^2_1(\lambda_i = x_{2i}^2)\), and the true ELPD over \(x_2\) is given by

\begin{equation*}
    \eta(x_{2}) = \E(\ell \mid x_{2})  =
    -\frac{1}{2}\log 4\pi - \frac{1}{4}(1 + x_{2}^2)\;.
\end{equation*}
Note that when $|x_2|$ is small, the distribution of $\ell$ is close to the central $\chi_1^2$ distribution with a heavy tail to the right generating extreme values. As $|x_2|$ grows, the distribution approaches the normal distribution but the variance grows large. This shows that modeling log scores is a hard problem.

\begin{figure}[htbp]
    \centering
    \makebox[\textwidth][c]{\includegraphics[width=1\textwidth]{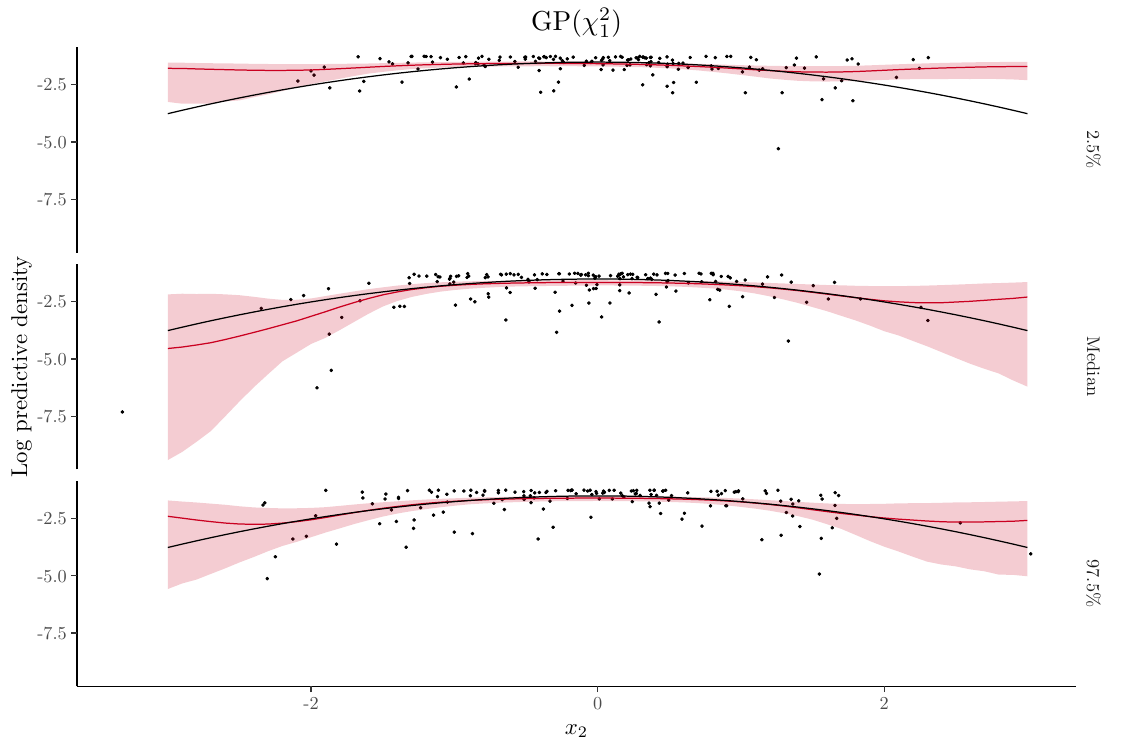}}
    \centering
    \makebox[\textwidth][c]{\includegraphics[width=1\textwidth]{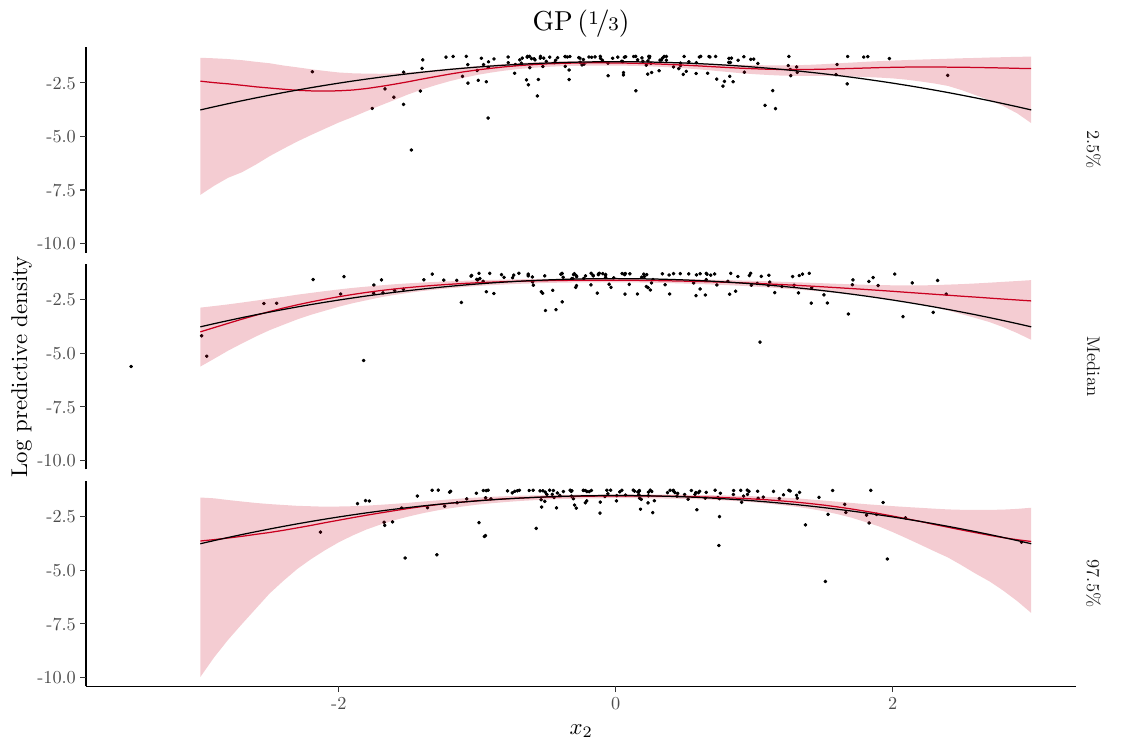}}
    \caption{Posterior distribution of \(\eta(x_{2})\) over a grid of \(x_2\) values for the \(\GP(\chi^2_1)\) (top) and \(\GP\left(\nicefrac{1}{3}\right)\) (bottom) models. The black line is the true \(\eta(x_{2})\). The red line is the posterior median and the red ribbon is the $95\%$ highest posterior density intervals. Median, \(2.5\%\), and \(97.5\%\) indicates which sample, out of $1000$, was used for each plot where the samples are ranked according to mean integrated squared error (MISE).}
    \label{sim_cubeGP_500}
\end{figure}


\begin{figure}[htbp]
    \centering
    \begin{subfigure}[b]{0.45\textwidth}
        \centering
        \includegraphics[width=\textwidth]{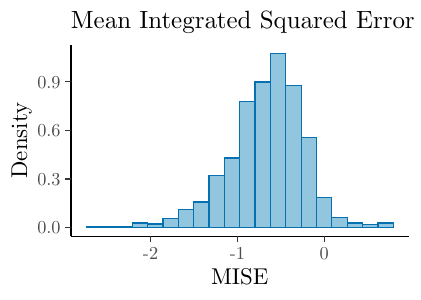}
        \label{fig:logmise}
    \end{subfigure}
    \hfill
    \begin{subfigure}[b]{0.45\textwidth}
        \centering
        \includegraphics[width=\textwidth]{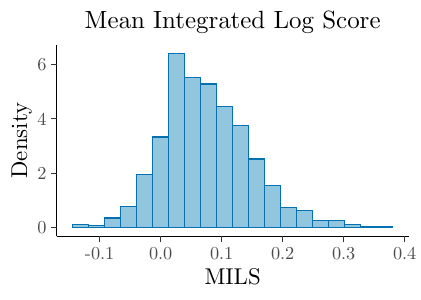}
        \label{fig:mils}
    \end{subfigure}
    \caption{The left figure shows the difference in log mean integrated squared error (MISE) of the \( \GP\left(\nicefrac{1}{3}\right)\) and \(\GP(\chi^2_1)\) model, taken over the distribution of \(x_2\). The right figure shows the difference in mean integrated log score (MILS) of the \( \GP(\nicefrac{1}{3})\) and \(\GP(\chi^2_1)\) model, taken over the distribution of \(x_2\).}
    \label{mise_mils}
\end{figure}

To compare the \(\GP\left(\chi^2_1\right)\) and \(\GP\left(\nicefrac{1}{3}\right)\) models we simulate $1 000$ data sets of $150$ observations each from the DGP in Eq. \eqref{simulation_dgp}. For each simulation, we then generate draws from the posterior distribution of \(\eta(x_{2})\) over a grid of \(x_{2}\) values between $-3$ and $3$, and compare with the true ELPD as a function of \(x_{2}\). Both GPs use a squared exponential kernel \citep{rasmussen_gaussian_2006}. We use \(l \sim \operatorname{Inverse-Gamma}(5,5)\) priors for the length scales, and a \(\alpha \sim \N^+(0, 1)\) prior for the signal variances. The noise variance \(\sigma_n^2\) of the \(\GP(\nicefrac{1}{3})\) is given a  \(\N^+(0, 1)\) prior as well.

Figure \ref{sim_cubeGP_500} show the posterior mean of \(\eta(x_{2})\) together with a \(95\%\) credible interval for three different data sets. The data sets have been selected by ranking the mean integrated squared error (MISE) of the ELPD estimates of each run---the MSE integrated over the distribution of $x_2$---and then selecting the \(2.5\):th, \(50\):th, and \(97.5\):th percentile. Both methods perform well for \(x_2 \in (-2, 2)\), but poorly in the tails. This is to be expected, as two factors conspire to make estimation in the tails difficult: the DGP variance of the log scores increases with the square of \(x_2\), and since \(x_{2} \sim \N(0, 1)\) only roughly $5\%$ of the observations are in the tails. 

Figure \ref{mise_mils} shows the difference in log MISE of the two models, and that the \(\GP\left(\nicefrac{1}{3}\right)\) model actually works marginally better in terms of squared error. However, the MISE metric only takes the posterior mean into account, so we also look at the mean integrated log score, which is calculated as the log predictive density of the true ELPD given the posterior distribution of \(\eta(x_2)\), integrated over the distribution of \(x_2\). The  \(\GP\left(\nicefrac{1}{3}\right)\) model works marginally better measured by this metric as well.

Note that the uncertainty around the estimate increases quickly for both models when there are no, or few, previous predictions nearby. This will lead to the desirable behavior that when combining experts in regions where we have little to go on in terms of previous predictions, we will quickly tend towards an equal-weights solution; see Sect. \ref{sec_empirical}.

The \(\GP(\chi^2_1)\) is prohibitively costly from a computational standpoint once the number of observations goes above a few hundred. That the more scalable \(\GP\left(\nicefrac{1}{3}\right)\) model works so well compared to the \(\GP(\chi^2_1)\) model is therefore an important finding for practical applications. There is also a large collection of methods for reducing computational cost in GP regression with additive homoscedastic errors \citep{liu_when_2020} that can be directly used to speed up inference in the \(\GP\left(\nicefrac{1}{3}\right)\) model.

\bibliographystyle{apalike}
\bibliography{my_lib}

\end{document}